\documentclass[preprint,prc,showpacs,preprintnumbers]{revtex4}
\usepackage{graphicx,latexsym,amssymb,amsmath,color,multirow,mathrsfs,booktabs,ifpdf}
\usepackage{dcolumn}
\usepackage{bm}
\usepackage{longtable}
\def\nA{nucleon-nucleus\ }
\def\oc{$^{16}$O+$^{12}$C\ }

\def\cc{$^{12}$C+$^{12}$C\ }
\def\oo{$^{16}$O+$^{16}$O\ }
\def\AA{nucleus-nucleus\ }
\begin{document}
\title{Nuclear mean field and double-folding model of the \AA 
optical potential}
\author{Dao T. Khoa$^1$}\email{khoa@vinatom.gov.vn} 
\author{Nguyen Hoang Phuc$^1$}
\author{Doan Thi Loan$^1$}
\author{Bui Minh Loc$^{1,2}$}
\affiliation{$^1$ Institute for Nuclear Science and Technology, VINATOM \\ 
179 Hoang Quoc Viet, Cau Giay, Hanoi, Vietnam. \\
$^2$ University of Pedagogy, Ho Chi Minh City, Vietnam}
\begin{abstract}
{Realistic density dependent CDM3Yn versions of the M3Y interaction have
been used in an extended Hartree-Fock (HF) calculation of nuclear matter (NM), 
with the nucleon single-particle potential determined from the total NM 
energy based on the Hugenholtz-van Hove theorem that gives rise naturally 
to a rearrangement term (RT). Using the RT of the single-nucleon potential 
obtained exactly at different NM densities, the density- and energy dependence 
of the CDM3Yn interactions was modified to account properly for both the RT and 
observed energy dependence of the nucleon optical potential. Based on a local 
density approximation, the double-folding model of the \AA optical potential 
has been extended to take into account consistently the rearrangement effect 
and energy dependence of the nuclear mean-field potential, using the modified 
CDM3Yn interactions. The extended double-folding model was applied to study the 
elastic \cc and \oc scattering at the refractive energies, where the Airy structure 
of the nuclear rainbow has been well established. The RT was found to affect 
significantly the real \AA optical potential at small internuclear distances, 
giving the potential strength close to that implied by the realistic 
optical model description of the Airy oscillation.}
\end{abstract}
\date{\today}
\pacs{24.10.Ht;25.70.Bc}
\maketitle

\section{Introduction}
\label{intro} 
During the last three decades, the double-folding model (DFM) of the \AA optical
potential (see, e.g., Refs.~\cite{Sa79,Kho95,Kho97,Kho00,Kho07r} and references therein) 
has been successfully used to calculate the real heavy-ion (HI) optical potential (OP) 
for use in different nuclear reaction studies. It is straightforward to see from the 
basic folding formulas that the folding model generates the first-order (Hartree-Fock type) 
term of the Feshbach's microscopic OP \cite{Fe92}. The success of the folding model 
description the observed elastic scattering of numerous HI systems, in particular, 
the nuclear rainbow pattern observed in the elastic scattering of the light HI
systems \cite{Br97}, clearly suggests that the first-order term of the Feshbach's 
microscopic OP is indeed the dominant part of the real \AA OP. 

The basic inputs for a folding model calculation are the nuclear densities of the 
colliding nuclei and the effective nucleon-nucleon (NN) interaction between the 
projectile nucleons and those in the target. A popular choice in the past for 
the effective NN interaction has been the M3Y interaction \cite{Sa79} which was 
designed to reproduce the G-matrix elements of the Reid \cite{Be77} and Paris 
\cite{An83} NN potentials in an oscillator basis. The original (density 
independent) M3Y interaction was used with some success in the folding model 
calculation of the real HI optical potential at low energies \cite{Sa79}, where the 
elastic scattering data are sensitive to the potential only at the surface, near 
the strong absorption radius $R_{\rm s.a.}$. Situation becomes different in cases 
of the refractive \AA scattering with the observation of the nuclear rainbow pattern
\cite{Kho07r}, where the elastic data measured at large angles were shown to be 
sensitive to the real OP over a wider radial range, down to small distances 
$R<R_{\rm s.a.}$. Here, the original M3Y interaction failed to give a good 
description of the data, and several realistic choices of the density dependence 
were included into the M3Y interaction \cite{Kho95,Kho97,Kho07r,Ko82,Kho93} to 
account for the reduction of the \emph{attractive} strength of the effective NN 
interaction at high densities of the nuclear medium, as the two nuclei closely 
approach and overlap with each other at small distances. 

The explicit density dependence of the M3Y interaction considered in the present
work was parametrized \cite{Kho95,Kho97,Kho93} to reproduce the saturation 
properties of symmetric nuclear matter (NM) in the standard Hartree-Fock (HF) 
calculation. To have a reliable density dependent interaction for use at different 
energies, the nucleon OP in NM obtained in the HF calculation \cite{Kho93,Kho95s}
(or the high-momentum part of the HF single-nucleon potential) was used to adjust 
the explicit energy dependence of the density dependent M3Y interaction against 
the observed energy dependence of the nucleon OP. However, the HF single-nucleon 
potential \cite{Kho93,Kho95s} is roughly equivalent to the single-particle 
potential of the Brueckner-Bethe theory \cite{Brown}, which lacks the 
\emph{rearrangement} term that arises naturally in the Landau theory for infinite 
Fermi systems \cite{Migdal}. Such a rearrangement term (RT) also appears when the 
single-nucleon potential is evaluated from the total NM energy using the Hugenholtz 
and van Hove (HvH) theorem \cite{HvH}, which is exact for all systems of interacting 
fermions, independent of the type of the interaction between fermions. 

For infinite NM, it is straightforward to see that the HvH theorem is satisfied 
on the HF level only when the in-medium NN interaction is density independent, 
i.e., when the RT is equal zero \cite{Cze02}. As a result, the single-nucleon 
(or nucleon mean-field) potential in NM evaluated on the HF level using an 
in-medium, density dependent NN interaction is not compliant  with the HvH 
theorem. It is of interest, therefore, to have a method to take into account 
properly the RT of the single-nucleon potential in NM on the HF level using 
the same density dependent NN interaction that was determined to reproduce the 
saturation properties of symmetric NM. Based on the exact expression of the RT 
of the single-nucleon potential given by the HvH theorem at each NM density 
and the empirical energy dependence of the nucleon OP observed over a wide range 
of energies, a compact method has been suggested recently \cite{Loa15} to account 
effectively for the RT in the standard HF scheme, by supplementing the density 
dependent CDM3Yn interaction \cite{Kho97} with the explicit contributions 
of the RT and of the momentum dependence of the nucleon mean-field potential. 

For finite nuclei, the RT appears naturally \cite{Bru59,Vau72} when the 
variational principle is applied to solve the eigenvalue problem in the HF 
calculation, using an effective density dependent NN interaction. Such a RT 
in the HF energy density of finite nuclei is known to describe the rearrangement 
of the mean field due to the removal or addition of a single particle 
\cite{Hee12}. In fact, it has been observed experimentally in the nucleon 
removal reactions at low energies that the interaction between the projectile 
nucleon and a target nucleon can induce some rearrangement of the 
single-particle configurations of other nucleons in the target \cite{Hs75}. 
In terms of the \AA interaction, such a rearrangement effect should also 
affect the shape and strength of the microscopic \AA OP constructed in the 
folding model using the single-particle wave functions of the projectile-
and target nucleons. Because the standard double-folding calculations of the 
\AA potential are being done mainly on the HF level \cite{Kho97,Kho00,Kho07r}, 
the impact of the rearrangement effect to the folded \AA OP has not been 
studied so far.  
 
The present work is the first attempt to address this important issue. For
this purpose, an extended version of the DFM is proposed to effectively include 
the RT into the folded \AA OP in the same mean-field manner, using 
consistently the same density- and momentum dependent CDM3Yn interaction 
determined from the extended HF study of NM \cite{Loa15} to be compliant 
with the HvH theorem. The extended DFM is applied to study in details 
the impact of the rearrangement effect of the nuclear mean field in the 
optical model analysis of the elastic, refractive \cc and \oc scattering at 
different energies. 

\section{Single-nucleon potential in the extended HF formalism}
\label{sec1}
We recall here the (nonrelativistic) Hartree-Fock description of homogeneous and 
symmetric NM at the given nucleon density $\rho$. Given the direct ($v^{\rm D}_{\rm c}$) 
and exchange ($v^{\rm EX}_{\rm c}$) parts of the (central) effective NN interaction 
$v_{\rm c}$, the ground-state energy of NM is $E=E_{\rm kin}+E_{\rm pot}$, 
where the kinetic and potential energies are determined as 
\begin{eqnarray}
 E_{\rm kin}(\rho) &=& \sum_{k \sigma \tau} n(k)
 \frac{\hbar^{2}k^2}{2m_\tau}  \label{ek1} \\
 E_{\rm pot}(\rho) &=& {\frac{1}{ 2}}\sum_{k \sigma \tau} \sum_{k'\sigma '\tau '}
 n(k) n(k')[\langle{\bm{k}\sigma\tau,\bm{k}'\sigma'\tau'}|v^{\rm D}_{\rm c}|
{\bm{k}\sigma\tau,\bm{k}'\sigma'\tau'}\rangle \nonumber \\
& & \hskip 2cm +\ \langle{\bm{k}\sigma\tau,\bm{k}'\sigma'\tau'}
|v^{\rm EX}_{\rm c}|{\bm{k}'\sigma\tau,\bm{k}\sigma'\tau'}\rangle] \nonumber \\
&=& {\frac{1}{ 2}}\sum_{k \sigma \tau} \sum_{k'\sigma '\tau '}
 n(k)n(k')\langle{\bm{k}\sigma\tau,\bm{k}'\sigma'\tau'}|v_{\rm c}|
{\bm{k}\sigma\tau,\bm{k}'\sigma'\tau'}\rangle_\mathcal{A}. \label{ek2} 
\end{eqnarray}
Here the single-particle wave function $|\bm{k}\sigma\tau\rangle$ is plane wave. 
The nucleon momentum distribution $n(k)$ in the spin-saturated, symmetric NM is 
a step function determined with the Fermi momentum $k_F=(1.5\pi^2\rho)^{1/3}$ as  
\begin{eqnarray}
 n(k)=\left\{\begin{array}{ccc}
 1 &\mbox{if $k \leqslant k_F$} \\
 0 &\mbox{otherwise.} \label{ek3}
\end{array} \right. 
\end{eqnarray}
According to the Landau theory for infinite Fermi systems \cite{Brown,Migdal}, 
the single-particle energy $\varepsilon(\rho,k)$ at the given nucleon density 
$\rho$ is determined as  
\begin{equation}
\varepsilon(\rho,k)=\frac{\partial E}{\partial n(k)}=t(k)+U(\rho,k)
=\frac{\hbar^2 k^2}{2m}+U(\rho,k), \label{ek4}
\end{equation}
which is the change of the NM energy caused by the removal or addition of 
a nucleon with the momentum $k$. The single-particle potential $U(\rho,k)$ 
consists of both the HF and rearrangement terms 
\begin{eqnarray}
U(\rho,k)&=& U_{\rm HF}(\rho,k)+U_{\rm RT}(\rho,k), \label{uk} \\
{\rm where}\ \ U_{\rm HF}(\rho,k) &=&\sum_{k'\sigma' \tau'}n(k')
\langle \bm{k}\sigma\tau,\bm{k}'\sigma'\tau'|v_{\rm c}|\bm{k}\sigma\tau,
\bm{k}'\sigma'\tau'\rangle_\mathcal{A} \label{uk1}\\
{\rm and}\ \ \ U_{\rm RT}(\rho,k) &=&\frac{1}{2}\sum_{k_1\sigma_1\tau_1}
 \sum_{k_2\sigma_2\tau_2}n(k_1)n(k_2) \nonumber \\
 &\times& \left\langle \bm{k}_1\sigma_1\tau_1,\bm{k}_2\sigma_2\tau_2\left|
 \frac{\partial v_{\rm c}}{\partial n(k)}\right|\bm{k}_1\sigma_1\tau_1,
 \bm{k}_2\sigma_2\tau_2\right\rangle_\mathcal{A}. \label{uk2}
\end{eqnarray}
It is clear from Eqs.~(\ref{ek4}) and (\ref{uk2}) that the RT accounts for the 
rearrangement of the nuclear mean field due to the removal or addition of a nucleon 
\cite{Hee12}. When the nucleon momentum approaches the Fermi momentum $(k\to k_F),\ 
\varepsilon(\rho,k_F)$ determined from Eqs.~(\ref{ek4})-(\ref{uk2}) is exactly the 
Fermi energy given by the Hugenholtz - van Hove theorem \cite{HvH}.  Using the 
transformation 
\begin{equation}
\frac{\partial }{\partial n(k)}\Bigg|_{k\to k_F}=
\frac{\partial \rho}{\partial n(k_F)}\frac{\partial k_F}{\partial \rho}
\frac{\partial }{\partial k_F}=\frac{1}{2\Omega}
\frac{\pi^2}{k_F^2}\frac{\partial }{\partial k_F}, 
 \label{ek5}
\end{equation}     
where $\Omega$ is the volume of symmetric NM, the RT of the single-particle 
potential $U$ at the Fermi momentum can be obtained as 
\begin{equation}
 U_{\rm RT}(\rho,k=k_F) = \frac{4\pi^2}{k_F^2} 
 \frac{\Omega}{(2\pi)^6}\int\!\!\!\!\int n(k_1)n(k_2)\left\langle\bm{k}_1\bm{k}_2
 \left|\frac{\partial v_{\rm c}}{\partial k_F}\right|
 \bm{k}_1\bm{k}_2\right\rangle_\mathcal{A}~d^3k_1 d^3k_2. \label{eRT}
\end{equation}
In difference from the RT part, the HF part of the single-particle potential 
is readily evaluated at any momentum 
\begin{equation}
 U_{\rm HF}(\rho,k) = \frac{4\Omega}{(2\pi)^3}\int n(k')\langle\bm{k}\bm{k}
|v_{\rm c}|\bm{k}\bm{k}'\rangle_\mathcal{A}~d^3k'. \label{eHF} 
\end{equation}    

For the spin-saturated symmetric NM, the spin and isospin components of plane 
waves are averaged out in the HF calculation of the single-particle potential, 
and only the spin- and isospin independent terms of the central NN interaction 
are needed for the determination of the single-particle potential 
(\ref{eRT})-(\ref{eHF}). In the present work, we use two density dependent 
versions (CDM3Y3 and CDM3Y6) \cite{Kho97} of the M3Y interaction based 
on the G-matrix elements of Paris potential in a oscillator basis \cite{An83}. 
Thus, the central CDM3Yn interaction is determined explicitly as
\begin{equation}
 v^{\rm D(EX)}_{\rm c}(s)=F_0(\rho)v^{\rm D(EX)}_{00}(s),\ {\rm where}\ 
 s=|\bm{r}_1-\bm{r}_2|. \label{CDM3Y}
\end{equation}  
The radial part of the interaction $v^{\rm D(EX)}_{00}(s)$ is kept unchanged 
as determined from the spin- and isospin independent part of the M3Y-Paris 
interaction \cite{An83}, in terms of three Yukawas
\begin{eqnarray}
v^{\rm D}_{00}(s)=11061.625\frac{\exp(-4s)}{4s}-2537.5\frac{\exp(-2.5s)}{2.5s}
+0.0002\frac{\exp(-0.7072s)}{0.7072s}, \nonumber \\
v^{\rm EX}_{00}(s)=-1524.25\frac{\exp(-4s)}{4s}-518.75\frac{\exp(-2.5s)}{2.5s}
 -7.8474\frac{\exp(-0.7072s)}{0.7072s}. \nonumber
\end{eqnarray}
The density dependence of the interaction (\ref{CDM3Y}) was assumed in 
Ref.~\cite{Kho97} as a hybrid of the exponential and power-law forms in order to 
obtain different values of the nuclear incompressibility $K$ in relatively small 
(10 to 20 MeV) steps from the HF calculation of NM. Given this empirical choice 
of $F_0(\rho)$, a realistic range for the $K$ value (the most vital input for
the equation of state of cold NM) has been deduced accurately from the folding 
model analysis of the refractive $\alpha$-nucleus and \AA scattering (see  
details in the review \cite{Kho07r}). Thus, we have used in the present 
work the following functional form for $F(\rho)$ \cite{Kho97}  
\begin{equation}
 F_0(\rho)=C[1+\alpha\exp(-\beta\rho)+\gamma\rho]. \label{fden}
\end{equation}
Note that the interaction (\ref{CDM3Y}) is the \emph{isoscalar} part of the
central interaction. A more comprehensive HF study of the nucleon mean-field 
potential in asymmetric NM has been performed recently \cite{Loa15}, taking 
into account also the isospin dependent part of the CDM3Yn interaction. In the 
DFM calculation of the \AA optical potential, the isospin dependent part of the 
effective NN interaction is needed only if both nuclei have nonzero isospins 
in their ground states \cite{Kho96}. In the present study we have focused, 
therefore, on the extension of the DFM using the spin- and isospin independent 
interaction (\ref{CDM3Y}) only.

\begin{figure}[bht] \vspace*{-1cm}
\includegraphics[width=0.9\textwidth]{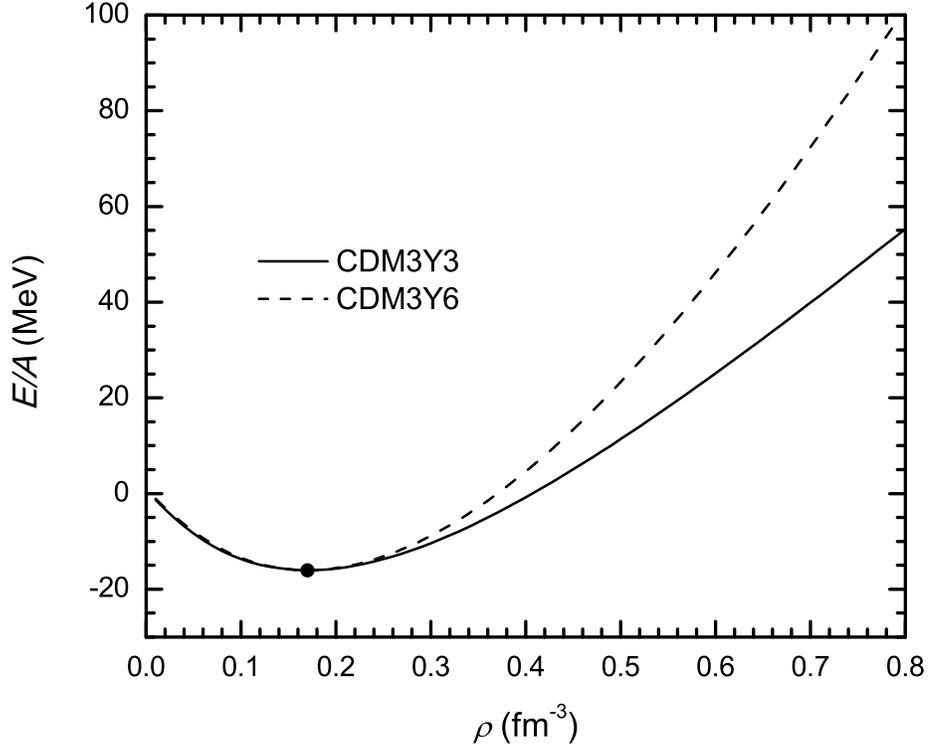}\vspace*{-0.5cm}
 \caption{Ground-state energy (per nucleon) of symmetric NM at different 
nucleon densities given by the HF calculation (\ref{ek1})-(\ref{ek2}), using the 
CDM3Y3 and CDM3Y6 interactions (\ref{CDM3Y}). The solid circle is the saturation 
point ($E/A\approx -15.9$ MeV at $\rho_0\approx 0.17$ fm$^{-3}$).} \label{f1}
\end{figure}
The parameters of the density dependence $F_0(\rho)$ were determined \cite{Kho97} 
to reproduce on the HF level the saturation properties of symmetric NM and 
give the nuclear incompressibility $K\approx 218$ and 252 MeV with the CDM3Y3 and 
CDM3Y6 version, respectively. These interactions, especially the CDM3Y6 version, 
have been widely tested in the folding model analyses of the elastic nucleus-nucleus 
scattering \cite{Kho97,Kho00,Kho07r}. The HF results for the ground-state energy 
of symmetric NM are shown in Fig.~\ref{f1}. One can see that at high NM densities 
the $E/A$ curve obtained with the CDM3Y6 interaction is stiffer than that obtained 
with the CDM3Y3 interaction, and this is due to the higher value of the nuclear 
incompressibility $K$ given by the CDM3Y6 interaction. 

\begin{table*}
\caption{Parameters of the density dependence $F_0(\rho)$ of the CDM3Yn 
interaction (\ref{fden}) and the correction $\Delta F_0(\rho)$ by the 
RT of the single-nucleon potential (\ref{fdelta}). The nuclear incompressibility 
$K$ is obtained in HF calculation of symmetric NM at the saturation density $\rho_0\approx 0.17$ 
fm$^{-3}$.} \label{t1} 
\begin{center}
\begin{tabular}{|c|c|c|c|c|c|c|} \hline
Interaction &   & $C$ & $\alpha$ & $\beta$ & $\gamma$ & $K$  \\
  &  &  &   & (fm$^3$) & (fm$^3$) & (MeV) \\ \hline
CDM3Y3 & $F_0(\rho)$  & 0.2985 & 3.4528 & 2.6388 & -1.5 & 218  \\
       & $\Delta F_0(\rho)$ & 0.38 & 1.0 & 4.484 & - & - \\ \hline
CDM3Y6 & $F_0(\rho)$ & 0.2658 & 3.8033 & 1.4099 & -4.0 & 252  \\
       & $\Delta F_0(\rho)$ & 1.50 & 1.0 & 0.833 & - & -  \\ \hline
\end{tabular}
\end{center}
\end{table*} 
Given the parametrization (\ref{CDM3Y}) of the central (isoscalar) NN interaction, 
the HF part of the single-nucleon potential can be explicitly obtained as
\begin{eqnarray}
 U_{\rm HF}(\rho,k) &=& F_0(\rho)U_{\rm M3Y}(\rho,k),\nonumber \\
 {\rm where}\ U_{\rm M3Y}(\rho,k) &=&\rho\left[J^D_0+
 \int\hat j_1(k_Fr)j_0(kr)v^{\rm EX}_{00}(r)d^3r\right]. \label{uHF} 
\end{eqnarray}
\begin{equation}
{\rm Here}\ J^{\rm D}_0=\int v_{00}^{\rm D}(r)d^3r, 
\ \hat{j_1}(x)= 3j_1(x)/x\ =\ 3(\sin x-x\cos x)/x^3. \nonumber 
\end{equation}
Applying the HvH theorem, the RT of the single-nucleon potential in symmetric
NM is obtained explicitly at the Fermi momentum as
\begin{equation}
U_{\rm RT}(\rho,k_F)=\frac{\rho^2}{2}\frac{\partial F_0(\rho)}{\partial\rho}
\left\{J^{\rm D}_0+\int [j_1(k_Fr)]^2 v_{00}^{\rm EX}(r)d^3r\right\}. \label{uRT}
\end{equation}
It is obvious from Eq.~(\ref{uRT}) that the RT becomes zero if the original
density-independent M3Y interaction is used in the HF calculation of the 
single-nucleon potential. In this case, the HvH theorem is satisfied already 
on the HF level \cite{Cze02}.

\begin{figure}[bht] \vspace*{-0.5cm}
\includegraphics[width=0.9\textwidth]{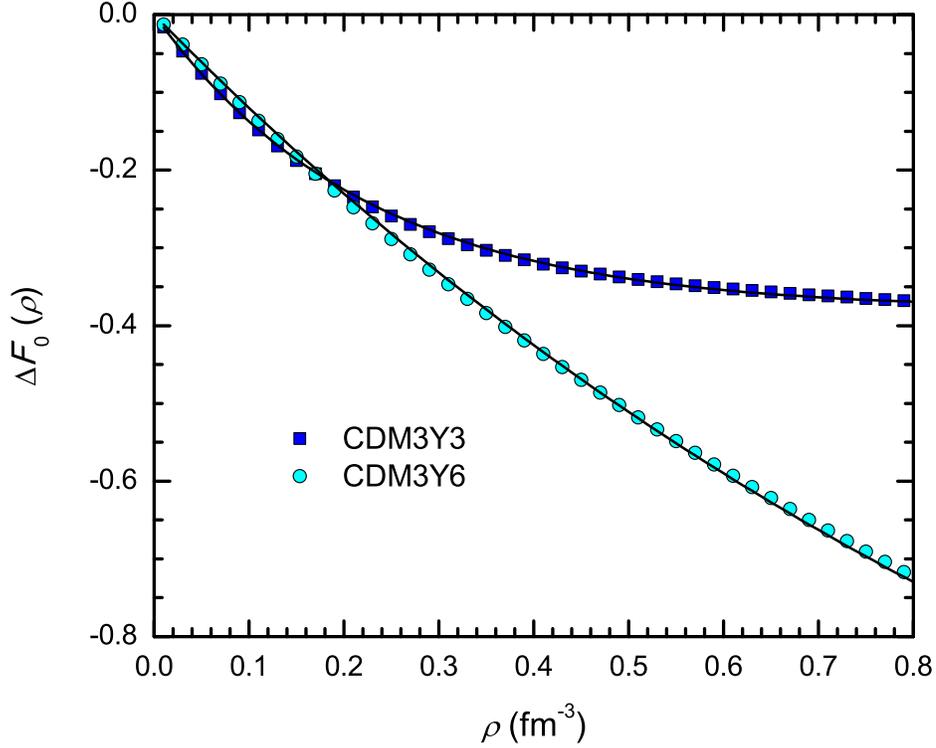}\vspace*{-0.5cm}
 \caption{Density dependence $\Delta F_0(\rho)$ of the RT (\ref{uRTk})
obtained from the exact expression of the RT given by the HvH theorem 
(\ref{uRT}) using Eq.~(\ref{dF}). Results obtained with the CDM3Y3 and CDM3Y6 
interactions are shown as squares and circles, respectively. The solid curves 
are given by the density dependent functional (\ref{fdelta}) using the 
parameters in Table~\ref{t1}.} \label{f2}
\end{figure}
In general, as seen from Eq.~(\ref{uk2}), the RT of the nucleon mean-field potential 
should be present at arbitrary nucleon momenta. Microscopically, the momentum 
dependence of the RT was shown, in the Brueckner-Hartree-Fock (BHF) calculations 
of NM \cite{Mah85,Zuo99}, to be due to the higher-order NN correlation, like the 
second-order diagram in the perturbative expansion of the mass operator or the 
contribution from three-body forces etc. In finite nuclei, the rearrangement 
effect in the nucleon removal reactions was shown \cite{Hs75} to be strongly dependent 
on the energy of the stripping reaction, a clear indication of the momentum dependence 
of the RT of the single-nucleon potential. Therefore, it is of interest to assess 
the momentum dependence of the RT of the single-nucleon potential on the HF level. 
Given the factorized density dependence of the CDM3Y3 and CDM3Y6 interactions, we 
have suggested recently \cite{Loa15} a compact method to account for the momentum 
dependence of the RT of the single-nucleon potential on the HF level. An important 
constraint for this procedure is that the momentum dependence of the total (HF+RT) 
single-nucleon potential reproduces the observed energy dependence of the 
nucleon OP. It was shown earlier \cite{Kho93,Kho95s} that the momentum dependence 
of the HF potential (\ref{uHF}) accounts fairly well for the observed energy 
dependence of the nucleon OP after a slight adjustment of the interaction strength 
at high energies. Therefore, in our extended HF formalism \cite{Loa15} the 
\emph{momentum dependent} RT of the single-nucleon potential is assumed to have 
a functional form similar to (\ref{uHF})   
\begin{equation}
U_{\rm RT}(\rho,k)= \Delta F_0(\rho)U_{\rm M3Y}(\rho,k), \label{uRTk}
\end{equation}
where the density dependent contribution of the rearrangement effect
is determined consistently from the exact expression (\ref{uRT}) of the RT 
at the Fermi momentum as   
\begin{equation}
 \Delta F_0(\rho)=\frac{U_{\rm RT}(\rho,k_F)}{U_{\rm M3Y}(\rho,k\to k_F)}. 
 \label{dF}
\end{equation}
Consequently, the total single-nucleon potential is determined in the extended 
HF approach as 
\begin{equation}
U(\rho,k)=[F_0(\rho)+\Delta F_0(\rho)]U_{\rm M3Y}(\rho,k). \label{Utotal}
\end{equation}
Thus, the momentum dependence of the total single-nucleon potential is 
determined by that of the exchange term of $U_{\rm M3Y}(\rho,k)$. One can see 
from the expressions (\ref{uRTk})-(\ref{Utotal}) that the rearrangement effect 
gives rise to a modified density dependence of the central interaction 
(\ref{CDM3Y}), $F_0(\rho)\to F_0(\rho)+\Delta F_0(\rho)$, and the total 
(HF+RT) single-nucleon potential (\ref{Utotal}) is readily obtained on the HF 
level. The density dependence of $\Delta F_0(\rho)$ obtained from the exact expression 
of the RT given by the HvH theorem at each NM density using Eq.~(\ref{dF}) is
shown as squares and circles in Fig.~\ref{f2}. One can see that the behavior 
of $\Delta F_0(\rho)$ at high NM densities is quite different for the two density 
dependent CDM3Yn interactions and this is associated with different NM 
incompressibilities $K$ given by these two interactions in the HF calculation of NM. 
Because $\Delta F_0(\rho)<0$ over the whole range of NM densities, the contribution
of the RT to the total single-nucleon potential is always repulsive.   
To facilitate the numerical calculation in the DFM, we have parametrized 
$\Delta F_0(\rho)$ using the a density dependent functional similar to (\ref{fden}) 
\begin{equation}
 \Delta F_0(\rho)=C[\alpha\exp(-\beta\rho)-1]. \label{fdelta}
\end{equation}
For convenience of the readers who are interested in using the modified CDM3Yn 
interaction (with the rearrangement contribution $\Delta F_0(\rho)$ added) in 
their folding model calculation, the parameters of $F_0(\rho)$ and 
$\Delta F_0(\rho)$ are given explicitly in Table~\ref{t1}. 

\begin{figure}[bht] \vspace*{-0.5cm}
\includegraphics[width=0.9\textwidth]{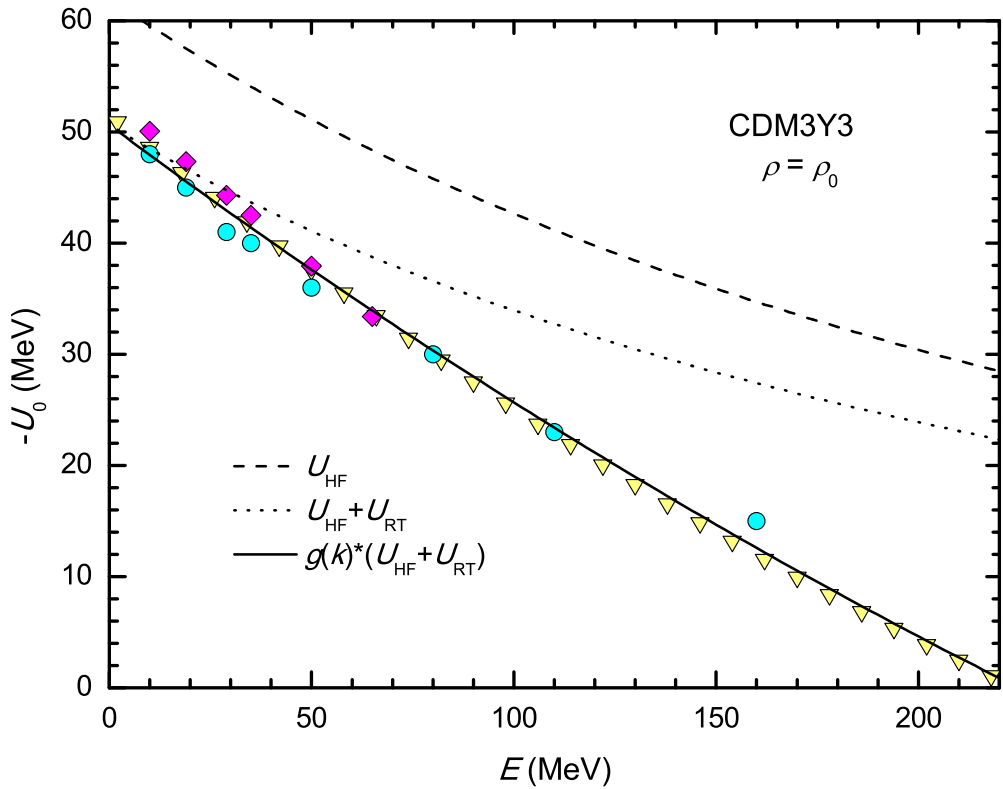}\vspace*{-0.5cm}
 \caption{Nucleon OP in symmetric NM, evaluated using the CDM3Y3 interaction
at the saturation density $\rho_0$ with and without the RT, in comparison with 
the empirical data taken from Refs.~\cite{BM69} (circles), \cite{Va91} (squares) 
and \cite{Hama} (triangles). The momentum dependent factor $g(k)$ was iteratively 
adjusted to the best agreement of the total nucleon OP (\ref{Uop0}) with the 
empirical data (solid line).} \label{f3}
\end{figure}
In the NM limit, the nucleon OP is determined as the mean-field interaction 
between the nucleon incident on NM at a given energy $E$ and bound nucleons in 
the filled Fermi sea \cite{Kho95s}. Applying a {\em continuous} choice for the
single-nucleon potential \cite{Ma91} at positive energies $E$, we obtain in the 
HF scheme the nucleon OP in \emph{symmetric} NM \cite{Kho93,Kho95s} as 
\begin{equation}
 U_0(\rho,E)=U_{\rm HF}(\rho,E)=F_0(\rho)\rho\left[J^D_0+
 \int\hat j_1(k_Fr)j_0\big(k(E,\rho)r\big)v^{\rm EX}_{00}(r)d^3r\right]. \label{Uop}
\end{equation}
Here $k(E,\rho)$ is the (energy dependent) momentum of the incident nucleon 
propagating in the mean field of bound nucleons, and is determined as
\begin{equation}
 k(E,\rho)=\sqrt{\frac{2m}{\hbar^2}[E-U_0(\rho,E)]},\ {\rm with}\ E>0. \label{Uopk}
\end{equation}
It is easy to see that $k(E,\rho)>k_F$ and $U_{\rm HF}$ is just the high 
momentum part of the HF potential (\ref{uHF}). Based on the above discussion, the 
total nucleon OP in the NM should also have a contribution from the RT added 
\begin{eqnarray}
 U_0(\rho,E)&=& U_{\rm HF}(\rho,E)+U_{\rm RT}(\rho,E) \nonumber\\ 
&=& [F_0(\rho)+\Delta F_0(\rho)]U_{\rm M3Y}\big(\rho,k(E,\rho)\big), \label{UopE} 
\end{eqnarray}
where the density dependence $\Delta F_0(\rho)$ of the RT is determined by the 
relation (\ref{dF}) and parametrized in Table~\ref{t1}. 
 
\begin{figure}[bht] \vspace*{-0.1cm}\hspace*{1cm}
\includegraphics[width=0.9\textwidth]{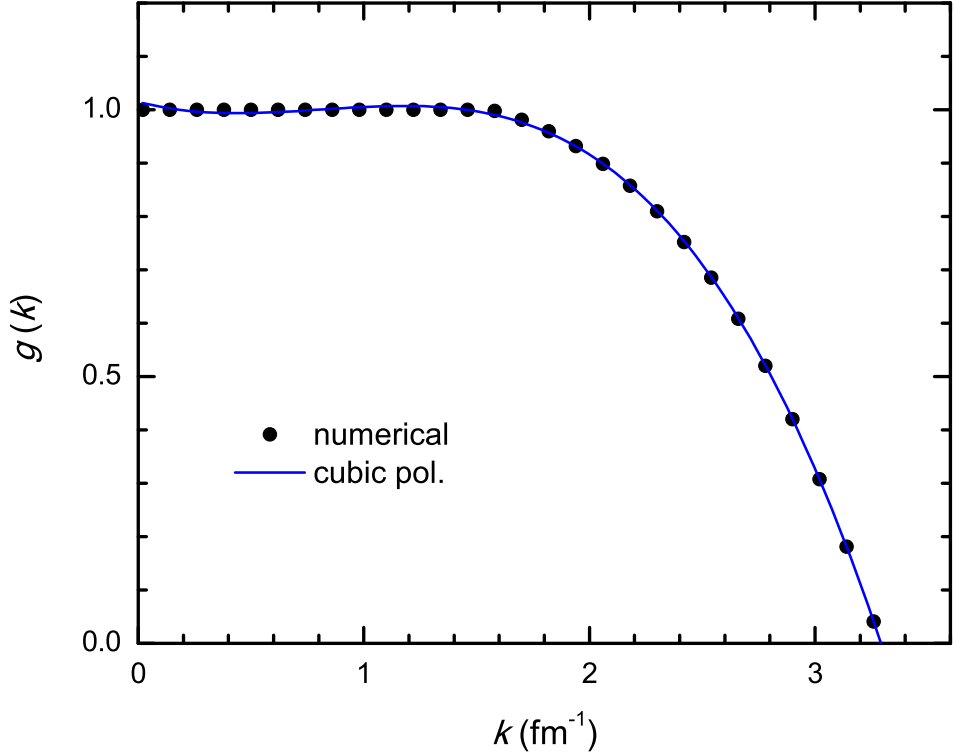}\vspace*{-0.5cm}
 \caption{Momentum dependent scaling factor $g(k)$ obtained with the CDM3Yn 
interaction from the best fit of the total nucleon OP (\ref{Uop0}) to the empirical 
data \cite{Hama} shown in Fig.~\ref{f3}. The points are the numerical results 
that are well reproduced by a cubic polynomial, $g(k)=1.015-0.109k+0.17k^2-0.07k^3$, 
(the solid line).} \label{f4}
\end{figure}
The total nucleon OP (\ref{UopE}) evaluated at the NM saturation density $\rho_0$ 
using the CDM3Y3 interaction is compared with the empirical data \cite{BM69,Va91,Hama} 
in Fig.~\ref{f3}. Although the inclusion of the RT significantly improves the 
agreement of the calculated $U_0$ with the data at lowest energies, it remains 
somewhat more attractive at high energies in comparison with the empirical 
data. This effect is easily understood in light of the microscopic BHF results for
the nucleon OP \cite{Ma91}, where the energy dependence of the nucleon OP in NM
was shown to originate from both the (energy dependent) direct and exchange parts 
of the Brueckner G-matrix. That is the reason why a slight linear energy dependence 
has been introduced into the CDM3Y6 interaction \cite{Kho97}, in terms of the 
$g(E)$ factor. To be consistent with the momentum dependence of the single-nucleon 
potential under study, instead of the $g(E)$ factor, we have introduced recently 
\cite{Loa15} a momentum (or energy) dependent scaling factor $g\big(k(E,\rho)\big)$ 
to the CDM3Yn interaction (\ref{CDM3Y}), and iteratively adjusted its strength to the 
best agreement of the (HF+RT) nucleon OP with the empirical data at the NM saturation 
density $\rho_0$ (see Fig.~\ref{f3}). Thus, 
\begin{equation}
U_0(\rho,E)=g\big(k(E,\rho)\big)[F_0(\rho)+\Delta F_0(\rho)]
U_{\rm M3Y}\big(\rho,k(E,\rho)\big), \label{Uop0} 
\end{equation}
where $k(E,\rho)$ is determined self-consistently from $U_0(E,\rho)$ via 
Eq.~(\ref{Uopk}). At variance with the $g(E)$ factor fixed by the incident 
energy \cite{Kho97}, $g\big(k(E,\rho)\big)$ scaling factor is a function of the 
(energy dependent) momentum of the incident nucleon (see Fig.~\ref{f4}), directly 
linked to the momentum dependence of the nucleon mean-field potential. 
Numerically, the obtained $g\big(k(E,\rho)\big)$ function is almost identical for 
both the CDM3Y3 and CDM3Y6 interactions and can be considered as the explicit 
momentum (or energy) dependence of the CDM3Yn interaction that allows the incident 
nucleon to feel the nucleon mean field during its interaction with nucleons bound 
in NM. Such a momentum dependence is of a similar nature as the momentum dependence 
of the G-matrix in the microscopic BHF study of NM, which is determined 
self-consistently through the momentum dependence of the single-particle energies 
embedded in the denominator of the Bethe-Goldstone equation \cite{Zuo99}. 
The technical difference here is that the $k$-dependence of $g(k)$ is determined 
empirically from the best fit of the calculated nucleon OP (\ref{Uop0}) 
to the observed energy dependence of the nucleon OP. Because $g(k)$ becomes 
smaller unity at $k\gtrsim 1.6$ fm$^{-1}$ (see Fig.~\ref{f4}), it has been used in our 
recent HF calculation \cite{Loa15} to adjust the high-momentum tail of the 
single-nucleon potential in NM. 

\section{Double-folding model of the \AA optical potential}
\label{sec2}
Given quite a strong rearrangement effect to the nucleon OP discussed above, 
it is of high importance to incorporate these effects in the many-body calculation 
of the \nA and \AA optical potentials. Based on the realistic treatment of the 
rearrangement effect and momentum dependence of the nucleon OP in the extended 
HF calculation of NM, a consistent inclusion of the RT into the single-folding 
calculation of the \nA OP for finite nuclei has been done \cite{Loa15} in the same 
mean-field manner, and the contribution of the RT was shown to be essential in 
obtaining the realistic shape and strength of the real nucleon OP. Because the 
double-folding model evaluates the \AA OP on the same HF-type level as the
single-folding calculation of the \nA OP, the contribution of the RT to the total
\AA potential is expected to be also significant. In the present work, we develop 
an extended version of the DFM to effectively include the RT into the double-folding
calculation of the \AA OP in a similar manner, using consistently the same 
density- and energy dependent CDM3Yn interaction that was fine tuned to be 
compliant with the HvH theorem in the HF study of NM discussed in Sec.~\ref{sec1}.  

We recall that in the DFM, the central \AA OP is evaluated as the HF-type 
potential \cite{Kho97,Kho07r} using an effective (energy- and density dependent) 
NN interaction $v_{\rm c}(\rho,E)$
\begin{equation}
  U_{\rm F}(E,R)=\sum_{i\in a,j\in A}[\langle ij|v^{\rm D}_{\rm c}(\rho,E)
	|ij\rangle +\langle ij |v^{\rm EX}_{\rm c}(\rho,E)|ji\rangle], \label{ef1}
\end{equation}
where $|i\rangle$ and $|j\rangle$ are the single-particle wave functions of the 
projectile ($a$) and target ($A$) nucleons, respectively. The direct part of the 
double-folded potential (\ref{ef1}) is local and expressed in terms of the 
ground-state (g.s.) densities of the two colliding nuclei as
\begin{equation}
 U_{\rm F}^{\rm D}(E,R)=\int\rho_a({\bm r}_a)\rho_A({\bm r}_A)
 v^{\rm D}_c(\rho,E,s)d^3r_a d^3r_A, \ \ {\bm s}={\bm r}_A-{\bm r}_a+{\bm R}.
\label{ef2}
\end{equation}
The antisymmetrization of the $a+A$ system is done by taking into 
account explicitly the knock-on exchange effects. As a result, the exchange term 
of $U_{\rm F}$ becomes nonlocal in the coordinate space \cite{Kho07r}. An accurate 
local approximation is usually made by treating the relative motion locally as 
a plane wave \cite{Kho07r}, and the exchange part of the double-folded potential 
(\ref{ef1}) can be obtained in the following \emph{localized} form 
\begin{eqnarray}
U_{\rm F}^{\rm EX}(E,R)&=&\int\rho_a({\bm r}_a,{\bm r}_a +{\bm s})
 \rho_A({\bm r}_A,{\bm r}_A -{\bm s}) \\ \nonumber
 &\times& v^{\rm EX}_c(\rho,E,s)\exp
\left(\frac{i{\bm K}(E,R).{\bm s}}{M}\right)d^3r_ad^3r_A, \label{ef3}
\end{eqnarray}
where $\rho_{a(A)}({\bm r},{\bm r}')$ are the nonlocal g.s. density matrices, 
$M=aA/(a+A)$ is the recoil factor (or reduced mass number), with $a$ and $A$ 
being the mass numbers of the projectile and target, respectively. The local 
momentum $K(E,R)$ of the relative motion is determined self-consistently from the 
real \AA OP as
\begin{equation}
 K^2(E,R)={{2\mu}\over{\hbar}^2}[E-U_{\rm F}(E,R)-V_C(R)], \label{ef4}
\end{equation}
where $\mu$ is the reduced mass of the two nuclei and $V_C(R)$ is the
Coulomb potential.

At low energies, the pair-wise interaction between the projectile nucleons  
and those in target can induce certain rearrangement of the single-particle 
configurations in both nuclei. Such impact by the rearrangement effect has been 
observed experimentally in the nucleon removal reactions \cite{Hs75}. In terms 
of the \AA interaction, the rearrangement effect should affect also the shape 
and strength of the \AA OP (\ref{ef1}), constructed in the DFM using the 
single-particle wave functions of the projectile- and target nucleons. Given 
the rearrangement contribution to the density dependence of the CDM3Yn interaction 
determined above in the HF study of the nucleon OP in NM, we have included the RT into 
the double-folding calculation of the \AA OP (\ref{ef1}) in the same manner as done
earlier for the single-folding calculation of the nucleon OP \cite{Loa15}. Namely,
the RT given by the correction $\Delta F_{0}(\rho)$ of the density dependence 
of the CDM3Yn interaction is added to the double-folded potential (\ref{ef1}), so that 
the total folded \AA OP is obtained in the present DFM calculation as
\begin{eqnarray}
 U_{\rm F}(E,R)&=& U_{\rm F}^{\rm D}(E,R)+U_{\rm F}^{\rm EX}(E,R)+U_{\rm RT}(E,R)
 \label{ef5} \\
  &=& g\big(k(E,R)\big)\int[F_0(\rho)+ \Delta F_{0}(\rho)] 
\big\{\rho_a({\bm r}_a)\rho_A({\bm r}_A)v^{\rm D}_{00}(s) \nonumber\\
  &+& \rho_a({\bm r}_a,{\bm r}_a +{\bm s})\rho_A({\bm r}_A,{\bm r}_A -{\bm s})  
 v^{\rm EX}_{00}(s)\exp\big(i{\bm k}(E,R).{\bm s}\big)\big\}d^3r_ad^3r_A, \nonumber
\end{eqnarray}
where the \emph{average} local nucleon momentum in the nuclear mean field of the two 
interacting nuclei is given by $k(E,R)=K(E,R)/M$. One can see in Eq.~(\ref{ef5}) 
that the contribution of the RT is present in both the direct and exchange terms 
of the \AA OP. Because the correction $\Delta F_0(\rho)$ by the RT is parametrized
in the density dependent form (\ref{fdelta})  similar to that of the CDM3Yn 
interaction (\ref{fden}), the double-folding integral (\ref{ef5}) can be readily 
evaluated using the DFM developed earlier in Refs.~\cite{Kho97,Kho00,Kho07r,Kho94}, 
where the sum of the two g.s. densities of the colliding nuclei is adopted for the 
overlap density $\rho$ appearing in $F_0(\rho)+\Delta F_0(\rho)$. Such a frozen density 
approximation (FDA) for the overlap density of the \AA system was discussed repeatedly 
in the past \cite{Sa79,Kho97,Kho07r,Kho94,Kho01}, and FDA was proven to be a reliable 
approximation at the (refractive) energies considered in the present study (see, e.g., 
results of the quantum molecular dynamics simulation of the \oo collision at 
22 MeV/nucleon \cite{Kho94} where the overlap density in the compression stage is very 
close to that given by the FDA, or the comparison of the FDA and adiabatic density 
approximation in Ref.~\cite{Kho01}). At low energies, especially, those of nuclear 
astrophysics interest, FDA is no more accurate and an appropriate adiabatic approximation 
for the overlap density should be used instead.    

It can be seen from Eqs.~(\ref{ef4})-(\ref{ef5}) that the energy dependence of the 
\AA OP folded with the CDM3Yn interaction is entirely determined by the local nucleon 
momentum $k(E,R)$ that appears explicitly in the exchange term as well as in the
local $g\big(k(E,R)\big)$ factor. Given the $g(k)$ function determined above 
in the HF calculation of NM based on the observed energy dependence of the nucleon 
OP, the local $g\big(k(E,R)\big)$ factor of the folded \AA potential (\ref{ef5}) 
is interpolated from the $g(k)$ values in the NM limit (see Fig.~\ref{f4}) for 
each local nucleon momentum $k=k(E,R)$ determined self-consistently from
Eq.~(\ref{ef4}) using an iterative procedure. Thus, $g\big(k(E,R)\big)$ can be 
considered as the explicit energy dependence of the density dependent CDM3Yn 
interaction (\ref{CDM3Y}), \emph{locally} consistent with the nuclear mean field 
based on the real folded \AA potential (\ref{ef5}). This is an important new feature 
of the extended DFM compared to the earlier versions of the DFM \cite{Kho97,Kho00} 
where a linear energy dependence $g(E)$ fixed by the incident energy was used to scale 
the CDM3Yn interaction. 

\section{Folding model analysis of the elastic \cc and \oc scattering at 
 the refractive energies}
\label{sec3}
In general, the elastic HI scattering is associated with the strong absorption
\cite{Sa79} that suppresses the refractive (rainbow-like) structure of the 
elastic cross section. Therefore, most of the elastic HI scattering occur 
at the surface and the measured elastic data carry little information 
about the \AA interaction potential at small distances ($R<R_{\rm s.a.}$). 
However, situation becomes different in cases of the refractive $\alpha$-nucleus 
or light HI scattering, where the absorption is weak and refractive structure 
of the nuclear rainbow appears at medium and large scattering angles, which enables 
the determination of the real \AA OP with a much less ambiguity, down to the 
sub-surface distances \cite{Kho07r}. The nuclear rainbow pattern has been shown 
to be of the far-side scattering, and is usually preceded in angles by the 
Airy minima \cite{Kho07r,Br97}. The observation of these minima, especially, 
the first Airy minimum A1 that is immediately followed by a broad (shoulder-like) 
nuclear rainbow pattern, greatly facilitates the determination of the real OP 
\cite{Kho07r,Fri88,Bra96}. It should be noted that the large-angle nuclear 
rainbow pattern observed in the (weak-absorbing) elastic $\alpha$-nucleus or 
light HI scattering can be shown, using the semi-classical formalism of the 
elastic \AA scattering developed by Brink and Takigawa 40 years ago \cite{Bri77}, 
to be associated with the \emph{internal} wave that penetrates 
through the Coulomb + centrifugal barrier into the interior of the real 
\AA OP, while the forward (diffractive) part of the elastic cross section is 
associated with the \emph{barrier} wave reflected from the barrier. 
As a result, the refractive (large-angle) elastic data are certainly sensitive 
to the real OP at small radii. In a further optical model (OM) study of the 
elastic \cc scattering at low energies, the broad (Airy-like) oscillation of the 
elastic cross section at medium and large angles was shown by Rowley {\it et al.} 
\cite{Row77} to be due to the interference of the far-side components of both the 
barrier and internal waves. Such an interference scenario is very similar (in the 
physics interpretation) to the Airy interference pattern of the nuclear rainbow 
discussed later in Refs.~\cite{Kho07r,Br97,Fri88,Bra96}.  

We recall here that there is a window in the energy or a range of the refractive 
energies, where the prominent nuclear rainbow pattern associated with the first 
Airy minimum can be observed. If the energy is too low, the broad rainbow pattern 
following A1 occurs at very backward angles and might well be hindered by other 
interferences (like the Mott interference in the symmetric \cc and \oo systems 
or the elastic $\alpha$-transfer in the \oc system). On the other hand, if the energy 
is too high, both A1 and the rainbow maximum move forward to small scattering angles 
and the rainbow structure is destroyed by the interference of the near-side and 
far-side scatterings, leading to the Fraunhofer oscillation. For the incident 
$^{12}$C and $^{16}$O ions, this energy window is about 10 to 40 MeV/nucleon, 
i.e., around the Fermi energy. It should be noted that the Airy interference pattern 
was also confirmed in the OM analyses \cite{Row77,McVoy92} of the elastic \cc scattering 
data at lower energies ($E<10$ MeV/nucleon) \cite{Wie76,Sto79}. In the present work, 
we concentrate mainly on the evolution of the broad nuclear rainbow pattern associated 
with A1, and the extended DFM is used to calculate the real OP for the OM analysis 
of the elastic \cc and \oc scattering data at the refractive energies.  

Like the earlier folding model studies of these data 
\cite{Kho97,Bra88,Kho94,Kho00O,Oglob00}, the (energy dependent) folded potential 
(\ref{ef5}) enters the OM calculation as the real OP and the imaginary part of the OP 
is assumed in the standard Woods-Saxon (WS) form. Thus, the total OP at the 
internuclear distance $R$ is determined as
\begin{equation}
 U(R)=N_RU_{\rm F}(E,R)-\frac{iW_V }{1+\exp[(R-R_V)/a_V]}+V_{\rm C}(R).
 \label{ef6} 
\end{equation}
The Coulomb part of the optical potential $V_{\rm C}(R)$ is obtained by directly 
folding two uniform charge distributions \cite{Pol76}, chosen to have RMS 
charge radii $R_C=3.17$ and 3.54 fm for $^{12}$C and $^{16}$O ions, 
respectively. Such a choice of the Coulomb potential was shown to be accurate up 
to small internuclear radii where the nuclear interaction becomes dominant 
\cite{Br97}. The g.s. nuclear densities of $^{16}$O and $^{12}$C used 
in the present DFM calculation were taken as Fermi distributions with parameters 
\cite{Kho01} chosen to reproduce the empirical matter radii of these nuclei.
All the OM analyses were made using the code ECIS97 written by Raynal \cite{Raynal}, 
with the relativistically corrected kinematics. The renormalization factor $N_R$ 
of the real folded potential and WS parameters were adjusted in each case to the best 
agreement of the calculated elastic cross section with the measured elastic data,
while keeping the shape of the complex OP within a consistent mean-field description.

\subsection{\cc system}\label{scc}
Among numerous experiments on elastic HI scattering, \cc is perhaps the most studied
system, with the elastic scattering measured at energies ranging from the Coulomb
barrier up to 200 MeV/nucleon. This is a strongly refractive system, with the
energy dependent Airy structure of the nuclear rainbow well established. The 
elastic \cc scattering data measured at different energies, over a wide angular 
range, enabled the determination of the real OP with a much less ambiguity. The deep
family of the real OP for this system (which is quite close to that predicted by
the folding model \cite{Br97}) gives a realistic evolution of the Airy minima  
shaping the famous ``Airy elephants" in the $90^\circ$ excitation function 
at low energies, where the prominent minimum at 102 MeV in the $90^\circ$ 
excitation function is due to the second Airy minimum A2 passing through 
$\theta_{\rm c.m.}\approx 90^\circ$ at that energy \cite{McVoy92}. The elastic \cc 
scattering was shown to be dominated by the far-side scattering at energies ranging 
from a few MeV/nucleon \cite{Row77,Bra88} up to 200 MeV/nucleon \cite{Hos88}. In 
the present work we consider selectively 6 data sets of the elastic \cc scattering 
measured at the incident energies of 139.5, 158.8, 240, 288.6, 360, and
1016 MeV \cite{Kubono83, Bohlen85, Alla10, Cole81, Buenerd84}, which were shown
in the earlier OM and folding model analyses \cite{Kho97,Br97,Bra88,Kho94,Kho00O} 
as sensitive to the strength and shape of the real OP at both the surface and 
sub-surface distances. In particular, the experiments on the \cc scattering at
$E_{\rm lab}=139.5$, 158.8 \cite{Kubono83} and 240 MeV \cite{Bohlen85, Alla10}
were aimed at revealing as clearly as possible the nuclear rainbow pattern. 

\begin{figure}[bht]\vspace*{-1.5cm}\hspace*{2cm}
\includegraphics[width=\textwidth]{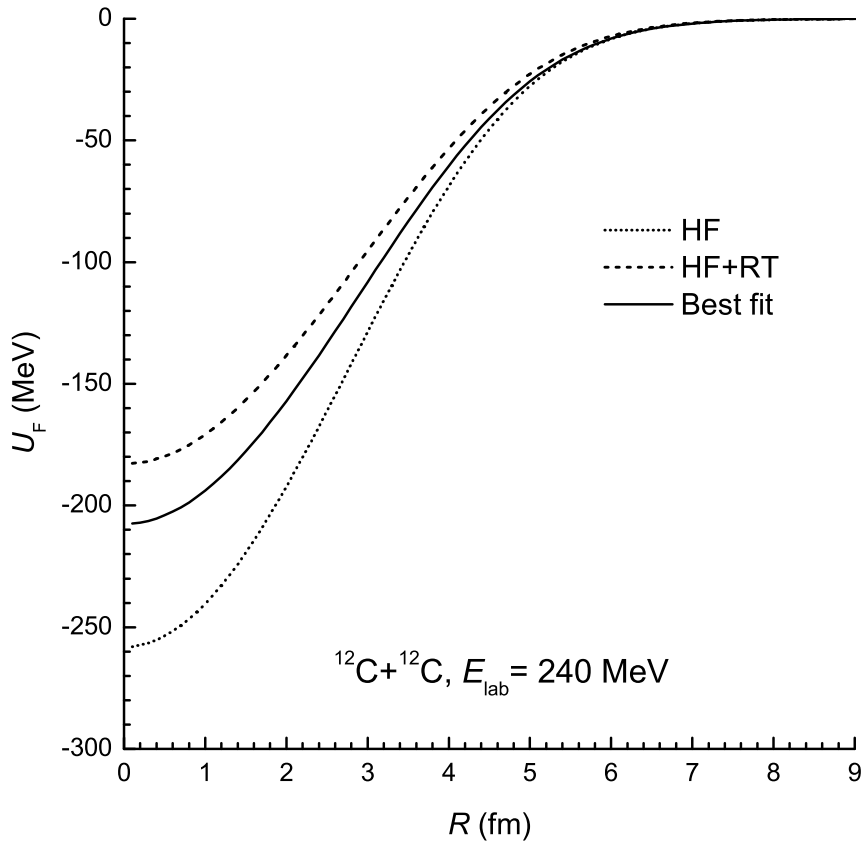}\vspace*{-3cm}
 \caption{Unrenormalized total (HF+RT) folded potential (\ref{ef5}) 
obtained with the CDM3Y3 interaction for the elastic \cc scattering at 
$E_{\rm lab}=240$ MeV (dashed line) in comparison with that obtained on the HF 
level only (dotted line), and the folded potential renormalized by the $N_R$ 
factor (solid line) determined from the best OM fit to the data (see 
Table~\ref{tcc}).} \label{fCC1}
\end{figure}
In the present work we focus on the impact of the rearrangement effect 
and momentum dependence of the nucleon mean-field potential in the folding
model description of the refractive \cc scattering. As shown earlier for the
folded nucleon OP, the rearrangement effect of the nucleon mean field gives
rise to a strong repulsive contribution of the RT to the real folded potential 
at small radii \cite{Loa15}. It is clear from Fig.~\ref{f2} that the higher 
the nuclear density the stronger the effect caused by the RT. In case of the 
double-folded \AA potential, the overlap nuclear density is well above $\rho_0$ 
at small internuclear distances and the repulsive contribution by the RT is quite
strong there. The (unrenormalized) total (HF+RT) real folded potential (\ref{ef5}) 
obtained with the CDM3Y3 interaction for the \cc system at $E_{\rm lab}=240$ MeV 
is compared with the HF folded potential in Fig.~\ref{fCC1} and one can see that 
the repulsive contribution of the RT is up to about $30\sim 40\%$ of the potential 
strength at the smallest radii. In the same direction, the best OM fit to the elastic 
\cc data at different energies requires also a shallower real OP compared to the deep 
HF folded potential which needs to be renormalized by a factor $N_R\approx 0.7\sim 0.8$ 
\begin{table}[bht]
\caption{The best-fit parameters of the OP (\ref{ef6}) for the elastic \cc 
scattering at $E_{\rm lab}=139.5-1016$ MeV. $N_R$ is the best-fit renormalization
factor of the real CDM3Y3 folded potential, $J_R$ and $J_W$ are the volume integrals 
(per interacting nucleon pair) of the real and imaginary parts of the OP, respectively.
$\sigma_R$ is the total reaction cross section.} 
\label{tcc} \vspace{0.5cm}
\begin{tabular}{|c|c|c|c|c|c|c|c|c|}\hline
 $E_{\rm lab}$ & Real OP & $N_R$ & $J_R$ & $W_V$ & $R_V$ & $a_V$ & $J_W$ & $\sigma_R$ \\
(MeV) &  &  & (MeV~fm$^3$) & (MeV) & (fm) & (fm) & (MeV~fm$^3$) & (mb) \\ \hline
139.5 &  HF & 0.810 & 325.5 & 26.60 & 5.170 & 0.600 & 121.2 & 1393  \\
  & HF+RT & 1.100 & 343.8 & 27.00 & 5.270 & 0.600 & 129.7 & 1443  \\ \hline 
158.8 & HF & 0.805 & 320.3 & 22.51 & 5.248 & 0.707 & 116.6 & 1514  \\
      & HF+RT & 1.101 & 337.7 & 23.25 & 5.290 & 0.740 & 119.5 & 1596 \\ \hline 
240  & HF & 0.815 & 311.9 & 24.02 & 5.425 & 0.645 & 127.1 & 1485  \\
     & HF+RT & 1.135 & 334.1 & 23.74 & 5.548 & 0.678 & 135.8 & 1596 \\ \hline
288.6  & HF & 0.805 & 300.7 & 26.73 & 5.122 & 0.715 & 124.6 & 1469  \\
      & HF+RT & 1.100 & 315.7 & 26.71 & 5.249 & 0.717 & 133.1 & 1519 \\ \hline
360  & HF & 0.693 & 250.3 & 22.91 & 5.180 & 0.672 & 108.0 & 1356 \\
     & HF+RT & 0.980 & 271.0 & 23.54 & 5.240 & 0.718 & 117.4 & 1455 \\ \hline 
1016 & HF & 0.670 & 183.8 & 17.09 & 5.171 & 0.802 & 85.03 & 1192  \\
     & HF+RT & 1.090 & 205.5 & 17.19 & 5.344 & 0.709 & 89.57 & 1187 \\ \hline  
\end{tabular}
\end{table}
for the best OM description of the data (see Table~\ref{tcc}). For the \cc system 
at the considered energies, the impact of the RT is slightly too repulsive and 
the total (HF+RT) folded potential needs to be renormalized by a factor 
$N_R\approx 1.1\sim 1.2$ for the best OM description of the data. We emphasize 
that the best-fit parameters of the OP presented in Table~\ref{tcc} allow one 
to properly describe the Airy structure of the nuclear rainbow pattern observed 
for the \cc system, using the CDM3Y3 folded potential. Using the real folded 
potential based on the CDM3Y6 interaction, we obtained the $N_R$ values about 
5\% larger than those obtained with the CDM3Y3 interaction and nearly the same 
WS parameters for the imaginary WS potential. Such a subtle effect is associated 
with a higher nuclear incompressibility $K$ given by the CDM3Y6 interaction in the 
HF calculation of NM, which results on a slightly more repulsive contribution 
of the RT to the total folded potential.   
\begin{figure}\vspace*{0cm}
\includegraphics[width=1.35\textwidth]{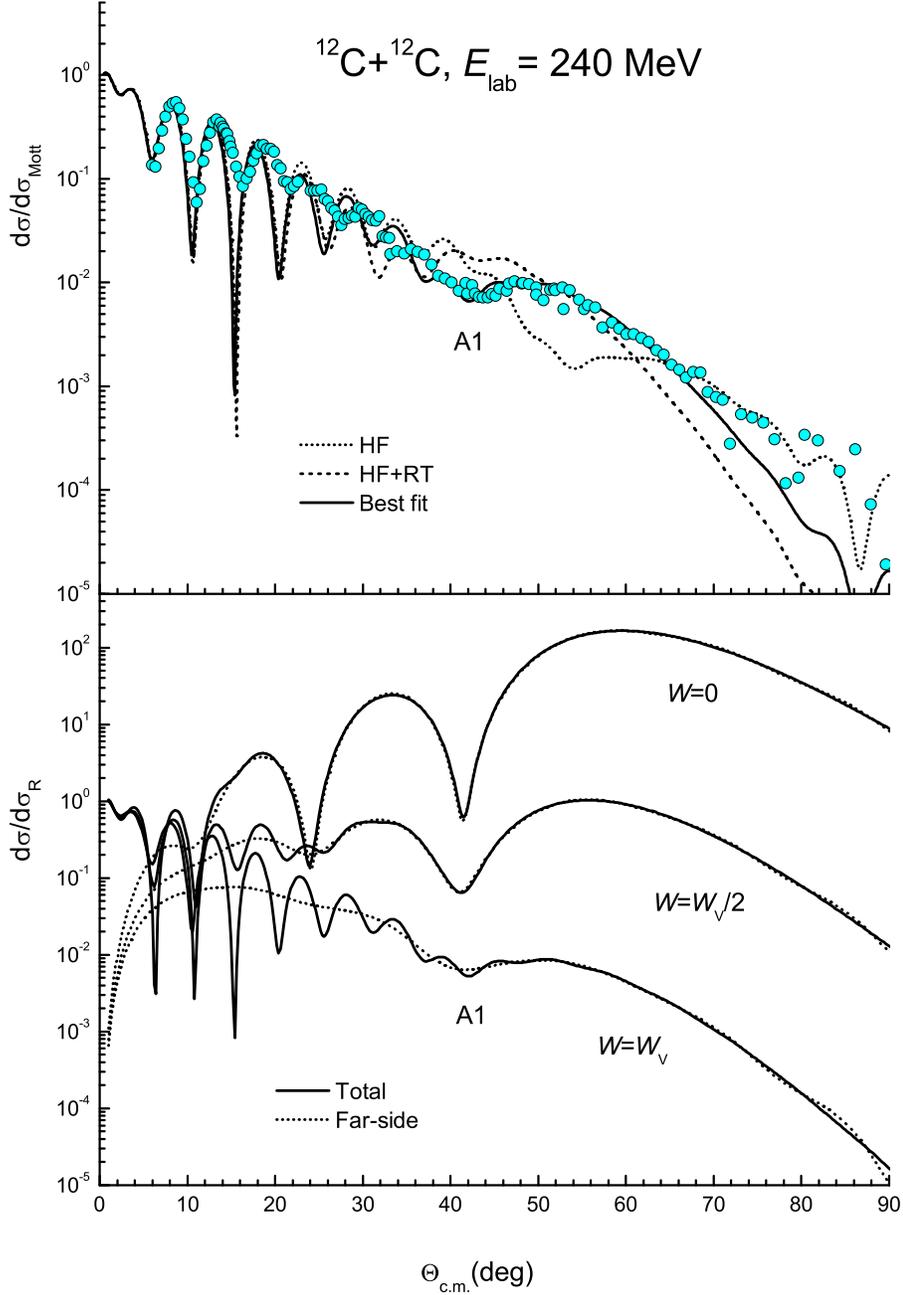}\vspace*{-4cm}
 \caption{Upper part: OM description of the elastic \cc scattering data at 
$E_{\rm lab}=240$ MeV \cite{Bohlen85,Alla10} given by three choices of the real 
folded potential (\ref{ef5}) shown in Fig.~\ref{fCC1}, using the best-fit imaginary 
OP taken from Table~\ref{tcc}. Lower part: Total unsymmetrized elastic \cc scattering 
cross section at 240 MeV (solid lines) and contribution of the far-side scattering 
(dotted lines) given by the best-fit real folded potential with different 
absorptive strengths of the WS imaginary potential taken from Table~\ref{tcc}.
A1 denotes the first Airy minimum that is followed by the broad (shoulder-like) 
rainbow pattern.} 
\label{fCC2}
\end{figure}

The elastic \cc scattering at 240 MeV is quite an interesting case (see Fig.~\ref{fCC2}). 
The first measurement of the elastic cross section at this energy \cite{Bohlen85} 
was done for angles up to $\theta_{\rm c.m.}\approx 55^\circ$ only, and the folding 
model analysis \cite{Bra88,Kho94} suggested that the first Airy minimum (A1) is 
located at $\theta_{\rm c.m.}\approx 41^\circ$ and followed by a broad (shoulder-like) 
rainbow pattern. However, an alternative scenario for the Airy structure in the 
\oc and \cc systems has been proposed \cite{Oglob03} where the minimum observed 
in the elastic \cc cross section at around $41^\circ$ is the second Airy minimum 
(A2), and the first Airy minimum A1 should occur at larger angles 
($\theta_{\rm c.m.}\approx 60^\circ$). To clarify the situation, a further 
experiment on the elastic \cc scattering at 240 MeV has been done using 
the kinematical coincidence method \cite{Alla10}. The elastic \cc scattering 
cross section measured up $\theta_{\rm c.m.}\approx 90^\circ$ (see upper part
of Fig.~\ref{fCC2}) show clearly no refractive minimum in the angular region
$\theta_{\rm c.m.}\approx 60^\circ\sim 70^\circ$. Therefore, A1 is now firmly
established at $\theta_{\rm c.m.}\approx 41^\circ$ for the elastic \cc scattering
at 240 MeV. As shown repeatedly in the earlier OM studies of elastic \cc
scattering \cite{Bra88,Kho94,McVoy92}, locations of the Airy minima are mainly 
determined by the strength of the real OP at small radii. One can see in 
Fig.~\ref{fCC1} that the (deeper) HF folded potential gives A1 located at 
$\theta_{\rm c.m.}\approx 54^\circ$ for the \cc system at 240 MeV, while A1 
given by the (shallower) HF+RT folded potential is shifted forward to around 
$38^\circ$. The best OM fit to these data given by the renormalized 
real folded potential and the WS imaginary potential (see Table~\ref{tcc})
reproduces the first Airy minimum around that observed in experiment at 
\begin{figure}[bht]\vspace*{-1.0cm}\hspace*{2cm}
\includegraphics[width=1.1\textwidth]{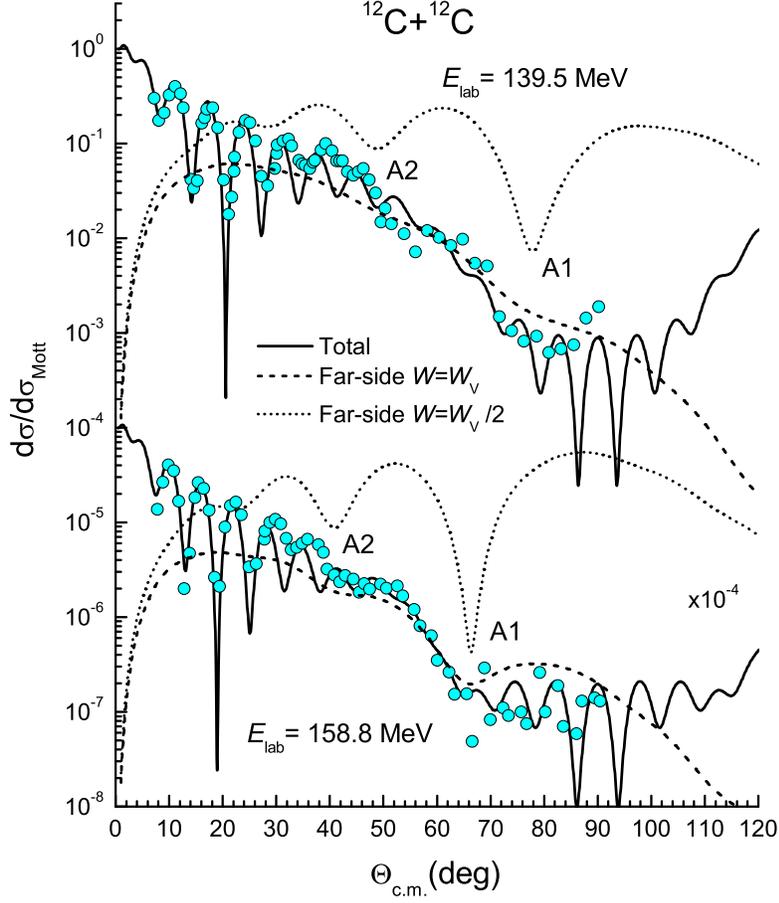}\vspace*{-4cm}
 \caption{OM description of the elastic \cc scattering data at 
$E_{\rm lab}=139.5$ and 158.8 MeV \cite{Kubono83} given by the best-fit (HF+RT) 
real folded potential and WS imaginary potential taken from Table~\ref{tcc} 
(solid lines). The far-side scattering cross sections are given by the 
unsymmetrized OM calculation using the same real folded OP but with different 
absorptive strengths $W_V$ of the WS imaginary potential (dashed and dotted 
lines). Ak is the k-th order Airy minimum} 
\label{fCC3}
\end{figure}
$\theta_{\rm c.m.}\approx 41^\circ$. Note that the best-fit elastic cross sections 
given by both the renormalized HF and HF+RT folded potentials are graphically 
the same as shown in upper part of Fig.~\ref{fCC2}. Although the Airy structure
of the nuclear rainbow is shaped by the real OP, its oscillating pattern is 
frequently obscured by the absorptive imaginary OP (or by the absorption of the 
incident flux into different nonelastic channels). To illustrate the refractive
(far-side scattering) structure of the elastic \cc scattering at this interesting
energy, we have performed the unsymmetrized OM calculation of the elastic
cross section (neglecting the Mott symmetrization required for the identical 
\cc system, to avoid the Mott oscillation of the elastic cross section at angles 
around $90^\circ$) using the best-fit HF+RT folded potential with different
absorptive strengths of the WS imaginary OP taken from Table~\ref{tcc}. 
The elastic cross section was further decomposed in terms of the near-side and 
far-side scattering cross sections using Fuller's method \cite{Ful75}, and one 
can see in the lower part of Fig.~\ref{fCC2} that the location of the first 
Airy minimum and the broad rainbow pattern that follows A1 are determined entirely 
by the far-side scattering amplitude, which in turn is determined mainly by the 
radial shape of the real OP. That's the reason why the accurate data of the nuclear 
rainbow scattering are indispensable in probing the strength and shape of the real 
\AA OP \cite{Kho07r,Br97}.
\begin{figure}[bht]\vspace*{-1cm}\hspace*{1cm}
\includegraphics[width=1.2\textwidth]{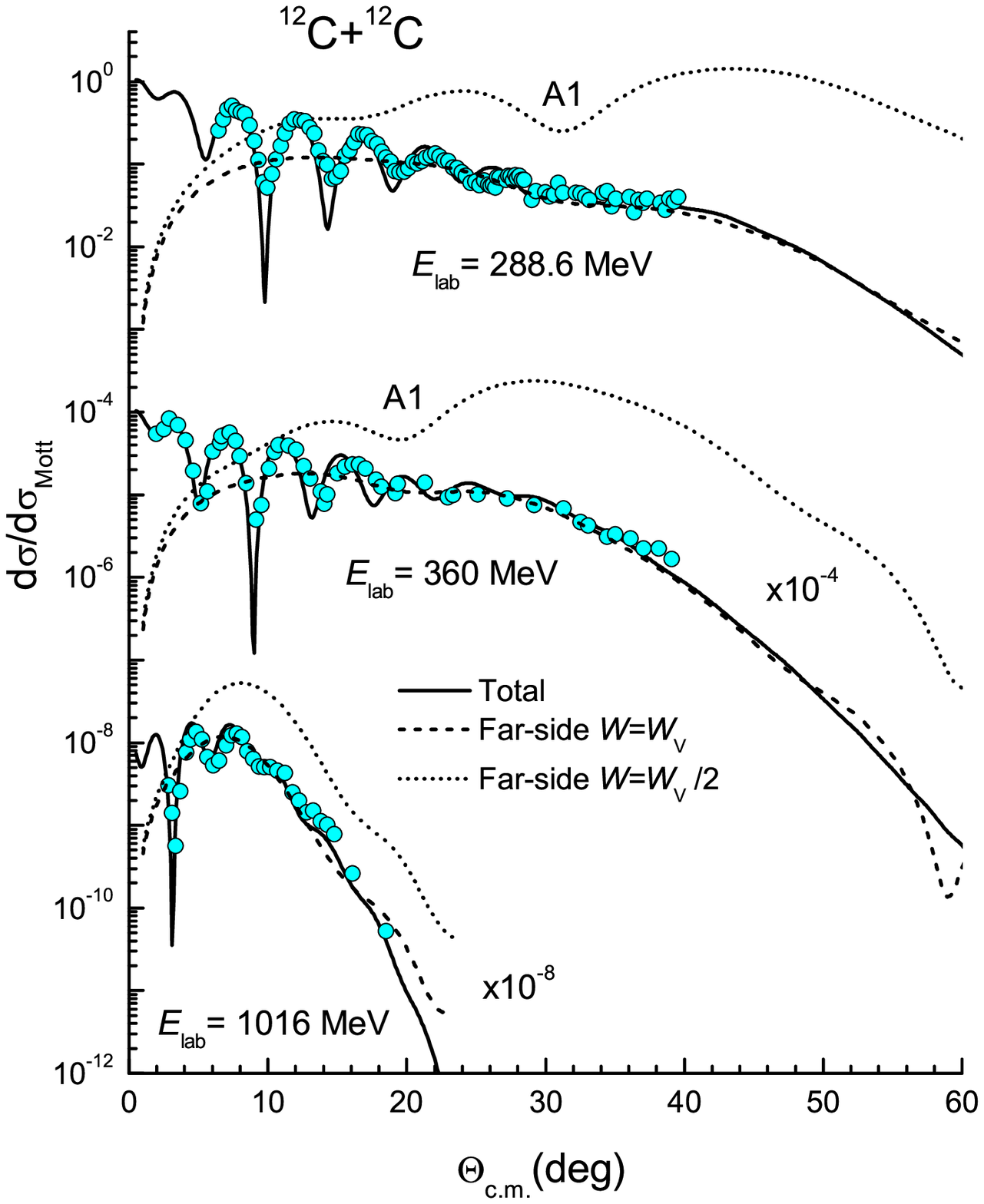}\vspace*{-4cm}
 \caption{The same as Fig.~\ref{fCC3} but for the elastic \cc scattering data 
measured at $E_{\rm lab}=288.6$, 360, and 1016 MeV \cite{Cole81, Buenerd84}.} 
\label{fCC4}
\end{figure}

At the lower incident energies of 139.5 and 158.8 MeV, because of the 
Mott interference at the scattering angles from $\theta_{\rm c.m.}\approx 70^\circ$ 
to beyond $90^\circ$, one cannot clearly allocate the Airy minima from the extensive data 
measured by Kubono {\it et al.} \cite{Kubono83}. Only the unsymmetrized OM calculation 
with different absorptive strengths could help to resolve that (see Fig.~\ref{fCC3}), 
and the first Airy minimum A1 was found at $\theta_{\rm c.m.}\approx 78^\circ$ and 
$66^\circ$ at the energies of 139.5 and 158.8 MeV, respectively.  
With the incident energies increasing to 288.6 and 360 MeV, A1 is shifted to the 
forward angles (see Fig.~\ref{fCC4}). A near-far decomposition of the scattering 
amplitude with a weaker absorption reveals the A1 location at 
$\theta_{\rm c.m.}\approx 31^\circ$ and $20^\circ$ at $E_{\rm lab}=288.6$ and 
360 MeV, respectively. One can see that the rainbow pattern associated with the 
first Airy minimum begins to be obscured by the near-far interference at 360 MeV. 
At the higher energy of $E_{\rm lab.}=1016$ MeV, the far-side scattering is still 
dominant at large angles, but becomes much weaker at the most forward angles where 
the scattering cross section shows a typical oscillation resulting from the
interference of the near-side and far-side scattering amplitudes. Given
a realistic (mean-field based) energy dependence of the CDM3Yn interaction
via $g\big(k(E,R)\big)$ factor [see Eqs.~(\ref{ef4})-(\ref{ef5})], the best-fit 
$N_R$ values obtained for the real HF+RT folded potential turned out to be around 
unity at the high energies of $E_{\rm lab.}=360$ and 1016 MeV, while those obtained 
for the real HF folded potential are below 0.7 (see Table~\ref{tcc}).   

Thus, we have shown that the evolution of the Airy structure in the elastic \cc 
scattering at the energies of 12 to 85 MeV/nucleon is well described by the real 
folded potential based on the modified density- and energy dependent CDM3Yn 
interaction that properly takes into account the rearrangement effect. With a 
strong impact of the RT to the nucleon mean-field potential at low energies 
\cite{Loa15}, the extended DFM should be further used in the OM study of the 
elastic \cc scattering at low energies, to pin down the potential ambiguity in 
the low-energy regime and improve the consistent mean-field description of the 
elastic scattering and shape resonances in the \cc system \cite{Kondo98}.

\subsection{\oc system}\label{soc}
Although the \cc system was shown above as strongly refractive, the Mott 
interference caused by the boson symmetry between the two identical 
$^{12}$C nuclei leads to rapidly oscillating elastic cross section at angles 
around $\theta_{\rm c.m.}=90^\circ$, which obscures the Airy structure in this 
angular region. As shown above in Figs.~\ref{fCC2}-\ref{fCC4}, the whole Airy 
pattern can be clearly seen only in the unsymmetrized OM calculation that removes 
the Mott interference artificially. The $^{16}$O+$^{12}$C system does not
have the boson symmetry, and was suggested 25 years ago by Brandan 
and Satchler \cite{Bra91} as a good candidate for the study of the nuclear 
rainbow. Since the late nineties, continuing efforts have been made by the 
Kurchatov-institute group to accurately measure the elastic \oc scattering
at the refractive energies ($E_{\rm lab}=132$ to 330 MeV) using the heavy-ion 
accelerators of both the Kurchatov institute and Jyv\"askyl\"a University
\cite{Oglob00,Oglob03,Oglob98,Demya}. In the present work we consider 
the elastic \oc scattering data measured at the incident energies of 132, 
170, 200, 230, 260, 281, and 330 MeV by the Kurchatov group \cite{Oglob00,Demya}
which exhibit quite prominent Airy structure of the nuclear rainbow, and
the elastic \oc scattering data measured at $E_{\rm lab}=300$ and 608 MeV 
by Brandan {\it et al.} \cite{Bra01,Bra86}. To study the energy dependence 
of the OP of the \oc system, the elastic data measured at 
$E_{\rm lab}=1503$ MeV \cite{Rou88} were also analysed in the present work. 

\begin{figure}[bht]\vspace*{-0.5cm}\hspace*{2cm}
\includegraphics[width=\textwidth]{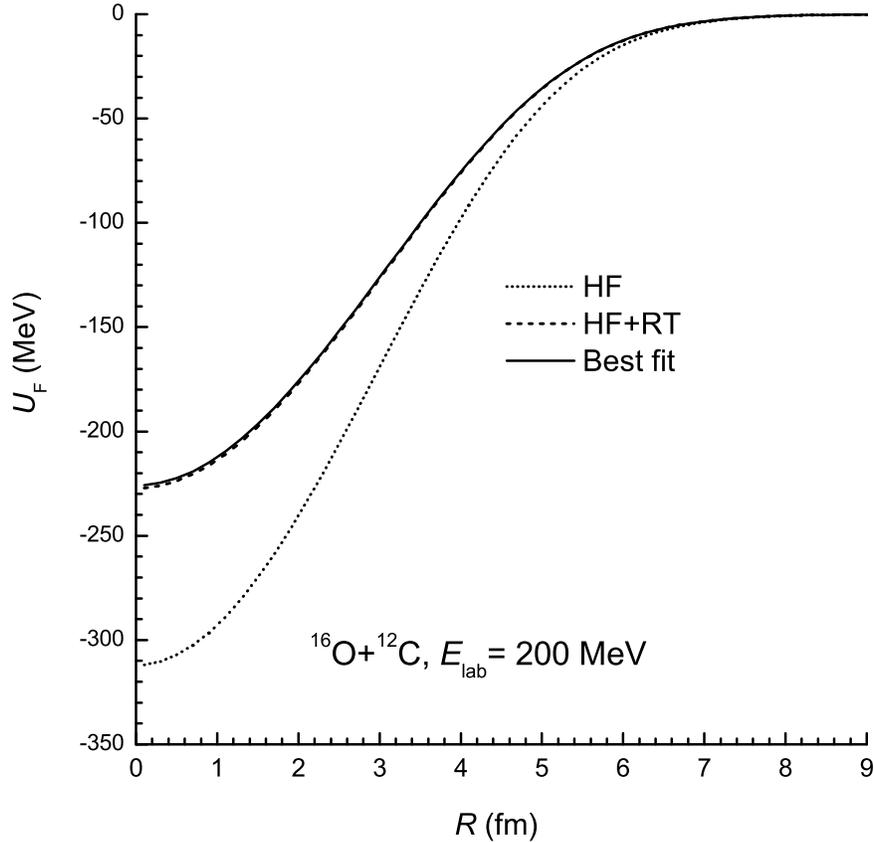}\vspace*{-3.5cm}
 \caption{Unrenormalized total real HF+RT folded potential (\ref{ef5}) obtained 
with the CDM3Y3 interaction for the elastic \oc scattering at $E_{\rm lab}=200$ 
MeV (dashed line) in comparison with that obtained on the HF level only (dotted 
line), and the real HF+RT folded potential renormalized by the best-fit $N_R$ 
factor taken from Table~\ref{toc} (solid line).} \label{fOC1}
\end{figure}
Among different elastic data measured at the refractive energies, the elastic 
\oc scattering data at $E_{\rm lab}=200$ MeV \cite{Oglob00} are perhaps the most 
prominent example of the nuclear rainbow observed in the light HI scattering. 
The (unrenormalized) HF and HF+RT real folded potential (\ref{ef5})  
obtained with the CDM3Y3 interaction for the \oc system at 200 MeV are shown 
in Fig.~\ref{fOC1}. As observed above for the \cc system, the repulsive 
contribution of the RT to the real folded \oc potential is up to about 40\% 
of the potential strength at the smallest radii. The best OM fit to the elastic 
\oc data at this energy also implies a real OP significantly shallower than 
the HF folded potential. It is remarkable that in the \oc case, the best-fit 
renormalization coefficient $N_R$ for the HF+RT folded potential is very close 
to unity, while that obtained for the HF potential is $N_R\approx 0.72\sim 0.75$ 
(see Table~\ref{toc}). This shows that the real folded potential obtained in the
extended DFM with the RT properly taken into account has a much improved 
predicting power for the real \AA OP, and the Airy structure of the elastic 
angular distribution observed for the \oc system at the considered energies 
can be reproduced rather well, using the unrenormalized real HF+RT folded potential. 
\begin{figure}\vspace*{-1.0cm}\hspace*{1cm}
\includegraphics[width=1.35\textwidth]{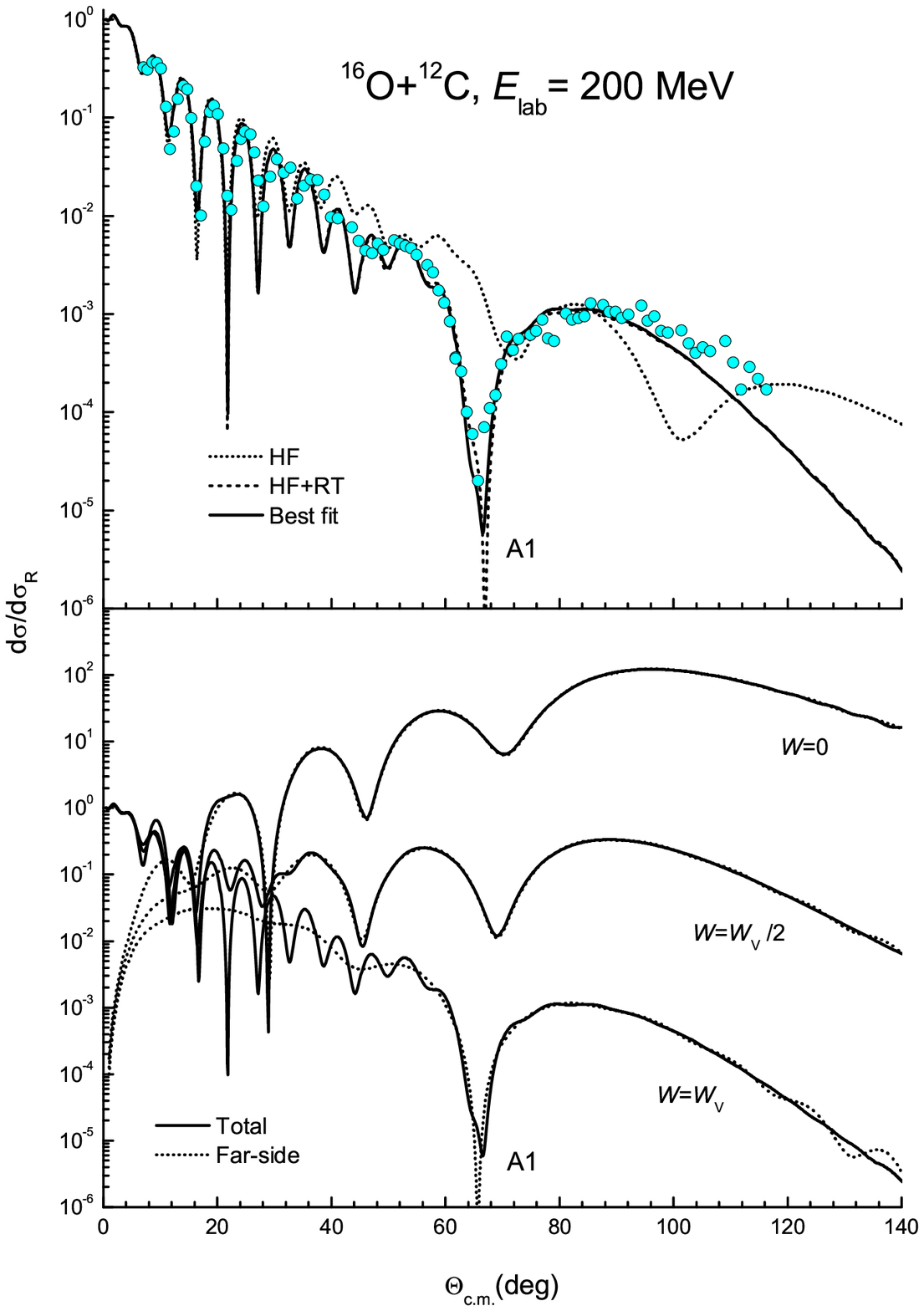}\vspace*{-3.5cm}
 \caption{Upper part: OM description of the elastic \oc scattering data at 
$E_{\rm lab}=200$ MeV \cite{Oglob00} given by three choices of the real 
folded potential (\ref{ef5}) shown in Fig.~\ref{fOC1}, using the best-fit imaginary 
OP taken from Table~\ref{toc}. Lower part: Total elastic \oc scattering 
cross section at 200 MeV (solid lines) and contribution of the far-side scattering 
(dotted lines) given by the best-fit real HF+RT folded  potential with different 
absorptive strengths of the WS imaginary potential taken from Table~\ref{toc}. 
A1 denotes the first Airy minimum that is followed by the broad (shoulder-like)
rainbow pattern.} \label{fOC2}
\end{figure}
The results of our folding model analysis of the elastic \oc scattering at 
$E_{\rm lab}=200$ MeV shown in Fig.~\ref{fOC2} illustrate very well the reliability 
of the HF+RT folded potential. Without the contribution of the RT, the HF folded 
potential is rather deep and wrongly predicts the first Airy minimum at 
$\theta_{\rm c.m.}\approx 101^\circ$ for the \oc system at 200 MeV. The shallower 
HF+RT folded potential shifts A1 forward to $\theta_{\rm c.m.}\approx 69^\circ$ 
as observed in the experiment. The best OM fit to these data resulted on the 
renormalization factor $N_R\approx 0.72$ and 0.99 for the HF and HF+RT folded
potentials, respectively. Not distorted by the Mott interference as in the \cc case,
the measured elastic \oc cross section at 200 MeV exhibits a broad shoulder-like 
rainbow pattern that spreads well over the angles beyond $100^\circ$. As a result, 
the elastic \oc data measured at this energy can serve as a very good probe of the 
real OP for the \oc system. Given much less ambiguity of the real OP in this case,
the strength and shape of the HF+RT folded potential (with the best-fit $N_R$ factor 
close to unity) turn out to be quite close to those implied by the realistic OM 
description of the Airy structure observed in the elastic \oc scattering data 
at 200 MeV \cite{Oglob00}. The strength of the HF folded potential needs to be 
scaled down by about 30\% to give a proper description of the first Airy minimum 
and the shoulder-like rainbow pattern that follows A1. Such a difference in the 
strength of the folded potential seems to be well accounted for by the repulsive 
contribution of the rearrangement term. 
\begin{figure}[bht]\vspace*{-1.0cm}\hspace*{1.5cm}
\includegraphics[width=1.1\textwidth]{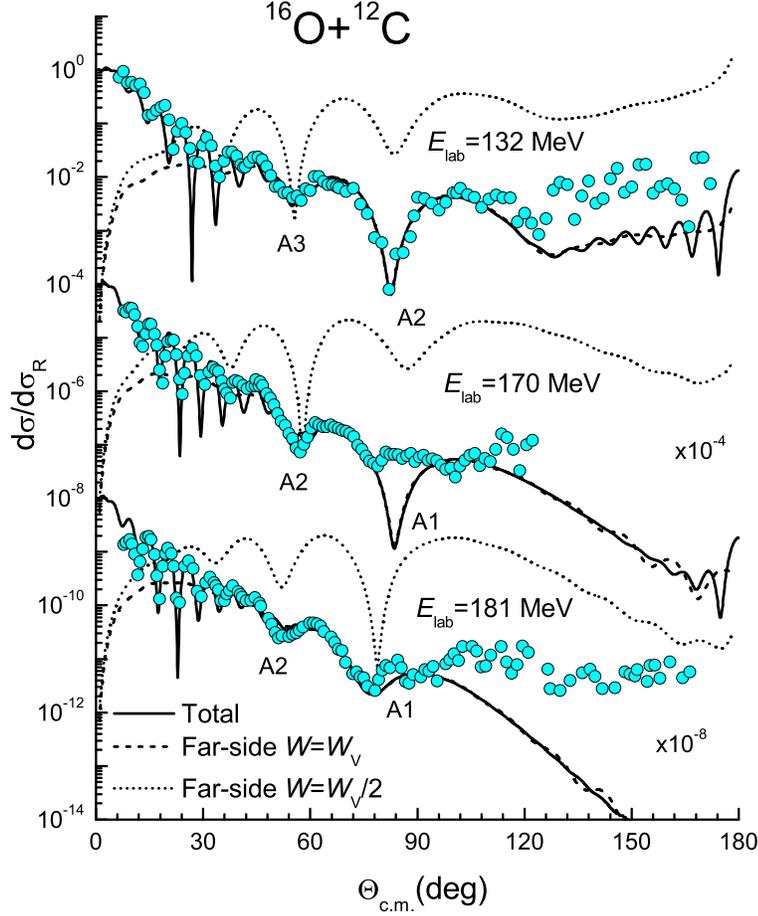}\vspace*{-4cm}
 \caption{OM description of the elastic \oc scattering data at 
$E_{\rm lab}=132$, 170 and 181 MeV \cite{Oglob00,Oglob03,Demya} given by the 
best-fit (HF+RT) real folded potential and WS imaginary potential taken from 
Table~\ref{toc} (solid lines). The far-side scattering cross sections are given by the 
same real folded OP but with different absorptive strengths $W_V$ of the WS imaginary 
potential (dashed and dotted lines). Ak is the k-th order Airy minimum.} \label{fOC3}
\end{figure}
The OM description of the elastic \oc scattering data at the lower energies of 
$E_{\rm lab}=132$, 170, and 181 MeV is shown in Fig.~\ref{fOC3}. The best fit to these
data has been achieved with the CDM3Y3 folded HF and HF+RT potentials renormalized 
by factor $N_R\approx 0.73-0.76$ and $N_R\approx 0.99-1.02$, respectively. This confirms 
again the important (repulsive) contribution of the RT to the \oc folded potential that 
helps to improve the prediction of the real OP by the extended DFM. The best-fit real 
folded potential also reproduces nicely the Airy oscillation established earlier
in the detailed OM analysis of these data \cite{Oglob00}. At 132 MeV, the most prominent
are the second (A2) and third (A3) Airy minima observed at $\theta_{\rm c.m.}\approx 
83^\circ$ and $56^\circ$, respectively. At this low energy, the first Airy minimum A1 
is located beyond $\theta_{\rm c.m.}=120^\circ$, and is totally obscured by the oscillating 
cross section at large angles that is likely due to the elastic $\alpha$ and nucleon
transfer processes \cite{vOe02,Mor11,Rud10}. The folded HF+RT potential could be used as 
the bare \oc potential in a future coupled reaction channel analysis of the elastic \oc
scattering at low energies, to study the contribution of the elastic $\alpha$-transfer
process.      

At 170 MeV, A1 and A2 are moved to $\theta_{\rm c.m.}\approx 87^\circ$ and $58^\circ$,
respectively. At 181 MeV, the locations of A1 and A2 are shifted to 
$\theta_{\rm c.m.}\approx 79^\circ$ and $52^\circ$, respectively. However, the  
primary rainbow pattern associated with A1 is still somewhat obscured and not clearly 
seen in the elastic data measured at 170 and 181 MeV. Thus, the most optimal energy 
for the observation of the primary rainbow pattern in the \oc system is 
$E_{\rm lab}=200$ MeV as shown in Fig.~\ref{fOC2}, and the measured data are 
a very valuable probe of the strength and shape of the real OP for this system 
as discussed in Fig.~\ref{fOC1}.   
\begin{table}\vspace{0cm}
\caption{The best-fit parameters of the OP (\ref{ef6}) for the elastic \oc 
scattering at $E_{\rm lab}=132-1503$ MeV. $N_R$ is the best-fit renormalization
factor of the real CDM3Y3 folded potential, $J_R$ and $J_W$ are the volume integrals 
(per interacting nucleon pair) of the real and imaginary parts of the OP, respectively.
$\sigma_R$ is the total reaction cross section.} 
\label{toc} \vspace{0.5cm}
\begin{tabular}{|c|c|c|c|c|c|c|c|c|}\hline
 $E_{\rm lab}$ & Real OP & $N_R$ & $J_R$ & $W_V$ & $R_V$ & $a_V$ & $J_W$ & $\sigma_R$ \\
(MeV) &  &  & (MeV~fm$^3$) & (MeV) & (fm) & (fm) & (MeV~fm$^3$) & (mb) \\ \hline
132 & HF & 0.757 & 316.5 & 13.31 & 5.937 & 0.642 & 67.81 & 1547   \\
  &  HF+RT & 1.017 & 331.7 & 14.67 & 5.772 & 0.751 & 71.80 & 1661  \\ \hline 
170 & HF & 0.741 & 305.3 & 17.69 & 5.913 & 0.579 & 87.80 & 1559  \\
  & HF+RT & 1.006 & 323.4 & 17.30 & 6.057 & 0.600 & 91.96 & 1624  \\ \hline 
181 & HF & 0.731 & 299.9 & 20.99 & 5.618 & 0.650 & 91.94 & 1561   \\
  & HF+RT & 0.987 & 315.9 & 20.91 & 5.733 & 0.650 & 96.87 & 1614 \\ \hline 
200 & HF & 0.723 & 294.5 & 18.26 & 5.959 & 0.550 & 91.41 & 1527  \\
  & HF+RT & 0.994 & 315.9 & 17.33 & 6.160 & 0.530 & 94.82 & 1579  \\ \hline
230 & HF & 0.726 & 292.4 & 21.02 & 5.776 & 0.597 & 97.71 & 1547  \\
  & HF+RT & 0.985 & 309.5 & 20.44 & 5.923 & 0.589 & 101.7 & 1597  \\  \hline
260 & HF & 0.716 & 285.2 & 21.99 & 5.740 & 0.555 & 99.11 & 1485  \\
  & HF+RT & 0.965 & 299.7 & 21.46 & 5.867 & 0.576 & 103.5 & 1563  \\  \hline
281 & HF & 0.707 & 279.4 & 22.58 & 5.685 & 0.552 & 98.93 & 1462 \\
  & HF+RT & 0.959 & 295.5 & 22.01 & 5.821 & 0.572 & 103.7 & 1542  \\  \hline	
300 & HF & 0.715 & 280.6 & 26.82 & 5.535 & 0.634 & 112.1 & 1550 \\
  & HF+RT & 0.960 & 293.6 & 26.37 & 5.630 & 0.680 & 117.4 & 1655  \\  \hline
330 & HF & 0.700 & 271.7 & 26.28 & 5.490 & 0.602 & 106.2 & 1476 \\
  & HF+RT & 0.945 & 285.7 & 24.99 & 5.653 & 0.600 & 109.4 & 1532  \\  \hline
608 & HF & 0.663 & 233.4 & 22.53 & 5.532 & 0.579 & 92.23 & 1359 \\
  & HF+RT & 0.915 & 247.7 & 21.48 & 5.745 & 0.586 & 97.96 & 1444  \\  \hline
1503 & HF & 0.671 & 179.5 & 19.01 & 5.511 & 0.758 & 82.38 & 1318 \\
  & HF+RT & 1.022 & 213.0 & 15.90 & 5.823 & 0.627 & 76.32 & 1262 \\  \hline
\end{tabular}
\end{table}

\begin{figure}[bht]\vspace*{-1.0cm}\hspace*{1.5cm}
\includegraphics[width=1.1\textwidth]{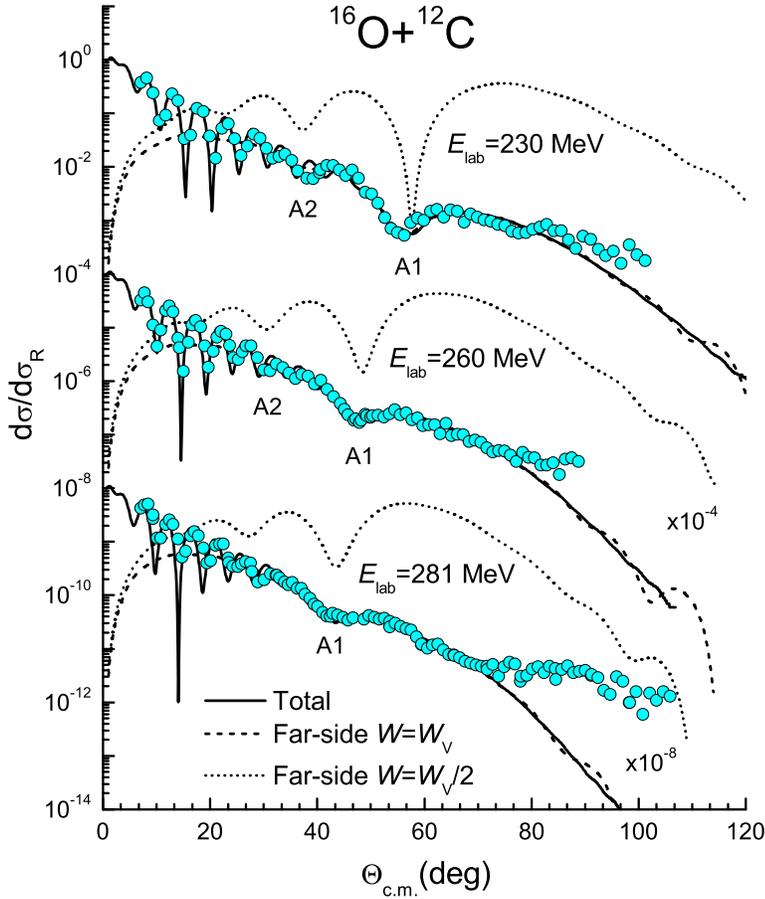}\vspace*{-4cm}
 \caption{The same as Fig.~\ref{fOC3}, but for the elastic \oc scattering data at 
$E_{\rm lab}=230$, 260, and 281 MeV \cite{Oglob00,Oglob03,Demya}.} 
\label{fOC4}
\end{figure}

\begin{figure}[bht]\vspace*{-1.0cm}\hspace*{1.5cm}
\includegraphics[width=1.1\textwidth]{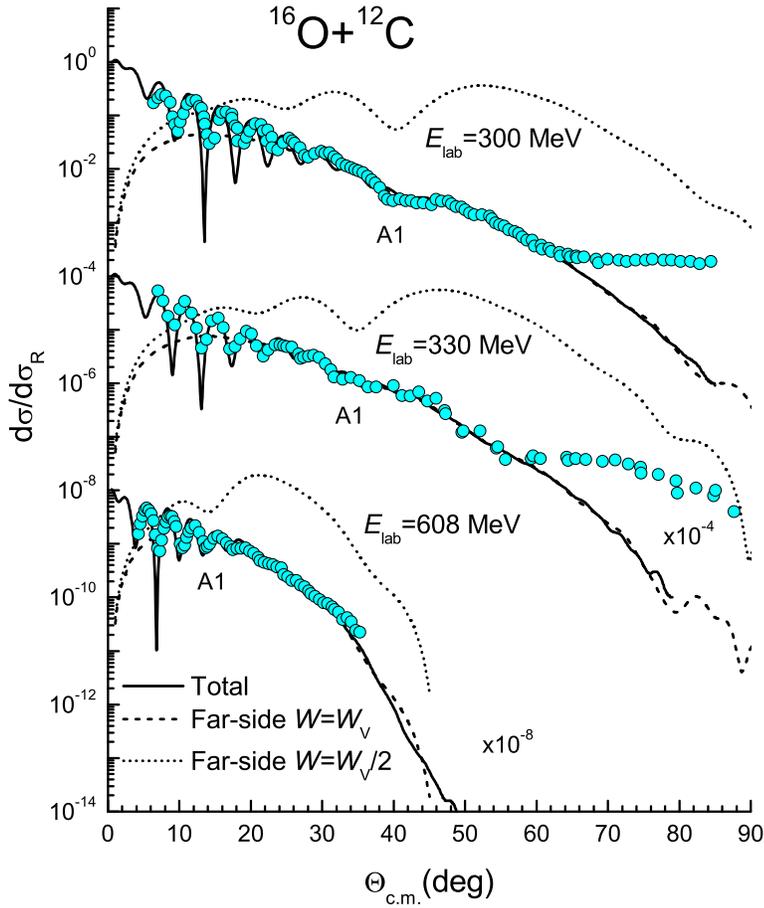}\vspace*{-4cm}
 \caption{The same as Fig.~\ref{fOC3}, but for the elastic \oc scattering data at 
$E_{\rm lab}=300$, 330, and 608 MeV \cite{Demya,Bra01,Bra86}.} 
\label{fOC5}
\end{figure}

With the incident energy increasing to above 200 MeV, the location of the Airy 
minima moves to the forward angles, as can be seen in the OM results for the 
elastic \oc scattering at $E_{\rm lab}=230$ to 608 MeV shown in 
Figs.~\ref{fOC4} and \ref{fOC5}. At 230 and 260 MeV, the first and second 
Airy minima are still visible in the measured data. The inspection of the
far-side scattering cross section, especially that with a weaker absorptive
strength of the imaginary OP, has shown that A1 is moved from 
$\theta_{\rm c.m.}\approx 58^\circ$ at 230 MeV to the c.m. angles of $48^\circ$
and $44^\circ$ at 260 and 281 MeV, respectively. Note that at  $E_{\rm lab}=281$
MeV the second Airy minimum A2 is moves into the diffractive part of the elastic 
cross section and is no more visible in the measured data.  
A distinct feature of the elastic \oc scattering data at $E_{\rm lab}=281$, 
300, and 330 MeV is the rise of the elastic cross section at large angles which is not
caused by the Airy interference of the far-side trajectories \cite{Kho07r,Br97}.
Because the shallow minimum associated with such a rise in the elastic cross section 
at large angles seems also moving slowly to smaller scattering angles with the
increasing energy, an alternative order of the Airy oscillation was suggested 
by Ogloblin {\it et al.} \cite{Oglob03} to accommodate one more shallow Airy minimum 
in the \oc system at large angles. However, such an Airy oscillation pattern could 
not consistently fit in the Airy structure established for this same \oc system 
at lower energies or other refractive systems like \cc as discussed 
above in Fig.~\ref{fCC2} or \oo \cite{Kho07r,Kho00O}. An interesting scenario 
for the backward rise of the elastic \oc cross section at these energies has been 
suggested recently by Ohkubo {\it et al.} \cite{Ohkubo14,Ohkubo14i} as due to 
a strong coupling of the elastic scattering to the inelastic 2$^+$ and 3$^-$ 
excitations of the $^{12}$C target. This important effect should be further 
checked by using other realistic choices of the OP and inelastic scattering potential 
for the \oc system. In particular, the elastic and inelastic folded potentials 
obtained in the present extended DFM that properly takes the RT into account 
should be a good choice for such a study. 
At the higher energies of $E_{\rm lab}=608$ and 1503 MeV the first Airy minimum 
A1 is well hidden in the forward angles, where only the Fraunhofer oscillation 
of the elastic cross section caused by the near-far interference 
\cite{Kho07r,Ful75} is visible in the measured data. Nevertheless, the large-angle 
exponential fall-off of the elastic cross section measured at these high energies 
is still dominated by the far-side scattering, and that has allowed us to determine 
the strength of the real folded (via $N_R$ renormalization factor) quite accurately.       
\begin{figure}[bht]\vspace*{0cm}\hspace*{1cm}
\includegraphics[width=1.0\textwidth]{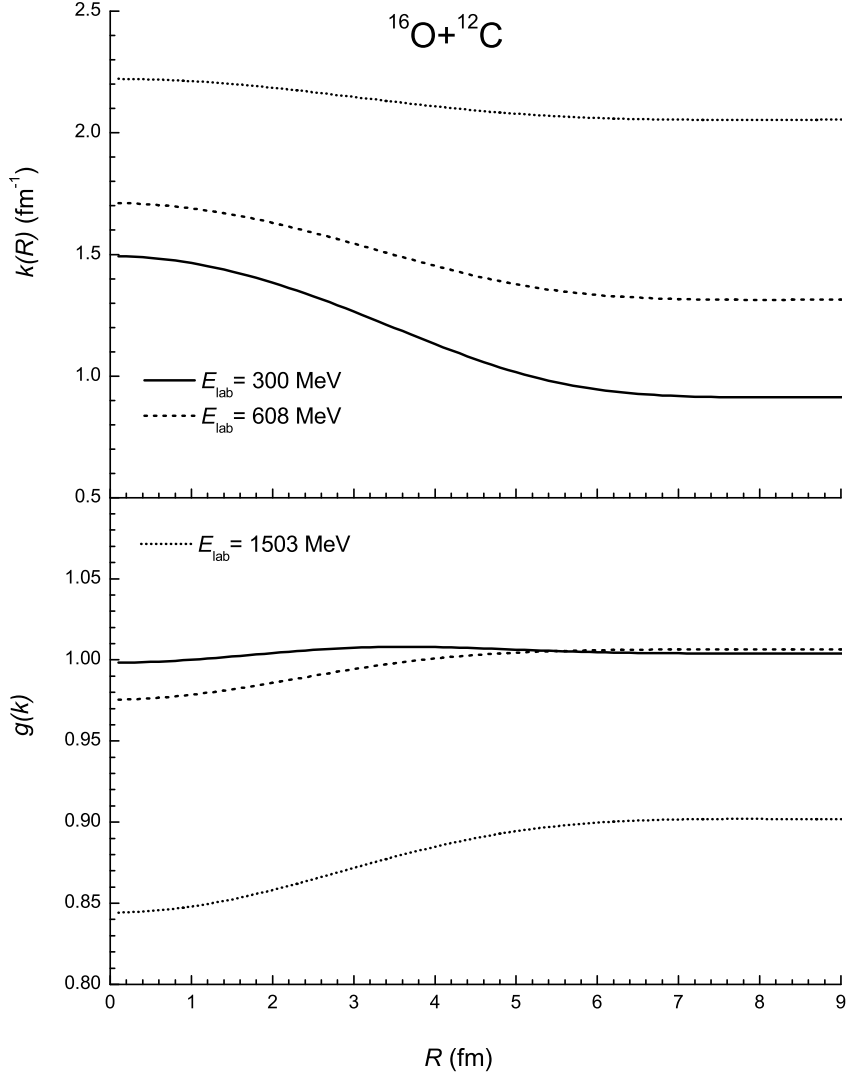}\vspace*{-1cm}
 \caption{Upper part: Local (average) nucleon momentum in the mean field 
based on the real folded \oc potential $k(E,R)=K(E,R)/M$, where $K(E,R)$ 
is the relative-motion momentum (\ref{ef4}). Lower part: Energy- and radial 
dependence $g\big(k(E,R)\big)$ of the CDM3Y3 interaction used in the folding 
calculation (\ref{ef5}) of the \oc potential, consistently interpolated from 
the explicit momentum dependence $g(k)$ of the nucleon OP shown in Fig.~\ref{f4}.} 
\label{fOC6}
\end{figure}

From the results of our detailed OM analysis of the elastic \oc scattering at 
energies up to 94 MeV/nucleon presented in Table~\ref{toc} one can see that the 
extended DFM accounts quite well for the energy dependence of the real OP. 
At energies up to 21 MeV/nucleon, the best fit to the elastic \oc data has been 
achieved with the CDM3Y3 folded HF potential renormalized by the factor 
$N_R\approx 0.70-0.75$, while the best-fit renormalization factor of the 
folded HF+RT potential is $N_R\approx 0.95-1.02$. However, at the higher energies
of 38 and 94 MeV/nucleon, the best-fit renormalization factor of the folded HF 
potential becomes $N_R\approx 0.66-0.67$. Given a realistic (mean-field based)
energy dependence of the CDM3Yn interaction in terms of the $g\big(k(E,R)\big)$ 
factor used in the extended DFM calculation (\ref{ef4})-(\ref{ef5}), the best-fit 
renormalization factor of the real folded HF+RT potential remains close to unity 
at high energies ($N_R\approx 0.92$ and 1.02 at the energies of 38 and 
94 MeV/nucleon, respectively). About the same trend was found with the best-fit 
$N_R$ factors given by the real CDM3Y6 folded potential, which are about 3-5\% 
larger than those obtained with the real CDM3Y3 folded potential. As discussed 
above in Sec.~\ref{scc}, this effect is associated with a higher nuclear 
incompressibility given by the CDM3Y6 interaction that leads to a more 
repulsion in the real folded potential at small radii. To illustrate the 
local energy dependence $g\big(k(E,R)\big)$ of the CDM3Y3 interaction used in 
the extended DFM calculation (\ref{ef5}) of the \oc potential,
we have plotted the local (energy dependent) nucleon momentum $k(E,R)$ and the 
corresponding $g\big(k(E,R)\big)$ factor in the upper and lower parts 
of Fig.~\ref{fOC6}, respectively. One can see that at low energies 
the real folded potential is deep at small distances $R$, so that the 
corresponding local relative-motion momentum $K(E,R)$ [see Eq.~(\ref{ef4})] or 
the average nucleon momentum $k(E,R)=K(E,R)/M$ is significantly higher 
at small $R$ compared to its asymptotic value at large $R$. The larger $K(E,R)$
value implies a quicker oscillation of the relative-motion wave function, and 
the deep real folded potential at low energies, as shown by Kondo {\it et al.} 
\cite{Kondo98}, usually generates the partial-wave radial functions having 
the numbers of nodes precisely as required by the Pauli principle when the 
dinuclear system is antisymmetrized. The local $g\big(k(E,R)\big)$ factor 
(see lower part of Fig.~\ref{fOC6}) was consistently interpolated from the 
explicit momentum dependence $g(k)$ of the nucleon OP in NM shown in 
Fig.~\ref{f4}, and it represents, therefore, the mean-field based energy 
dependence of the CDM3Yn interaction used in the extended DFM calculation 
of the \oc potential. At low energies ($E\lesssim 20$ MeV/nucleon) when 
$k(E,R)\lesssim 1.6$ fm$^{-1}$, the $g\big(k(E,R)\big)$ factors remain close 
to unity. At the high energy of 94 MeV/nucleon, the $k(E,R)$ value is increased 
to above 2 fm$^{-1}$ over the whole radial range (see upper part of Fig.~\ref{fOC6}) 
so that the corresponding $g\big(k(E,R)\big)$ factor is reduced significantly 
(see lower part of Fig.~\ref{fOC6}). Thus, the consistent treatment of the mean-field 
based energy dependence of the CDM3Yn interaction via $g\big(k(E,R)\big)$ factor 
helps to improve the predicting power of the real folded potential.

\section{Summary}
The CDM3Y3 and CDM3Y6 density dependent versions of the M3Y-Paris interaction
have been used in an extended HF study of symmetric NM, focusing on the 
rearrangement term of the single-nucleon potential that appears naturally
when the Hugenholtz-van Hove theorem is taken into account in the calculation
of the single-nucleon energy. Based on the exact expression of the RT of the 
density dependent single-nucleon potential given by the HvH theorem and the 
empirical energy dependence of the nucleon OP, a compact method has been proposed 
to account properly for the density- and energy dependence of the RT of the 
nucleon OP in NM on the HF level.   

Given an explicit contribution of the RT added to the density dependence 
of the CDM3Yn interaction and proper treatment of the momentum dependence 
of the nucleon mean-field potential in NM on the HF level, the double-folding
model has been extended to take into account consistently the rearrangement 
effect in the DFM calculation of the \AA OP in the same mean-field manner. 
The contribution of the RT to the total \AA folded potential has been shown to be 
repulsive and particularly strong at small internuclear distances. This result
is complimentary to the recent DFM calculation of the \AA OP using a G-matrix 
interaction that includes effectively the three-body force \cite{Minomo14}. 
In fact, the microscopic origin of the RT was shown in the BHF study of NM 
\cite{Mah85,Zuo99} to be due to the higher-order diagram in the perturbative 
expansion of the mass operator and the contribution of the three-body force.

The present extension of the DFM is an important milestone that allows us  
to obtain the realistic shape and strength of the real folded OP at small 
internuclear distances, which match closely those implied by the detailed OM 
analysis of the elastic \cc and \oc scattering data measured at the 
refractive energies. The realistic treatment of the (mean-field based) energy 
dependence of the CDM3Yn interaction in the extended DFM calculation significantly 
improves the predicting power of the real folded potential, especially, 
in the proper description of the Airy structure of the nuclear rainbow pattern
observed in the elastic \cc and \oc scattering. All parameters of the modified 
density dependence of the CDM3Yn interaction that takes into account the correction 
by the RT and the mean-field based energy dependence are given in such details 
that the interested readers could easily include these parameters into their 
folding model calculation. The present development of the DFM for the \AA OP 
can be generalized and applied further in the folding model study of the 
inelastic \AA scattering \cite{Kho00}, to reveal, in particular, the impact 
of the rearrangement effect caused by the nuclear excitation. This is the  
object of our further research.   

\section*{Acknowledgments}
The present research has been supported by the National Foundation for 
Scientific and Technological Development (NAFOSTED Project No. 103.04-2014.76).
D.T.K. is grateful to late Ray Satchler for his helpful discussions
on the mean-field description of the elastic \AA scattering. We also thank Alexei 
Ogloblin and Alla Dem'yanova for their communication on the elastic \oc scattering 
data.


\begin{thebibliography}{99}
\bibitem{Sa79} G.R. Satchler and W.G. Love, Phys. Rep. {\bf 55}, 183 (1979).
\bibitem{Kho95} D.T. Khoa and W. von Oertzen, Phys. Lett. B {\bf 342}, 6 (1995).
\bibitem{Kho97} D.T. Khoa, G.R. Satchler, and W. von Oertzen,
 Phys. Rev. C {\bf 56}, 954 (1997).
\bibitem{Kho00} D.T. Khoa and G.R. Satchler, Nucl. Phys. A {\bf 668}, 3 (2000).
\bibitem{Kho07r} D.T. Khoa, W. von Oertzen, H.G. Bohlen, and S. Ohkubo,
 J. Phys. G {\bf 34}, R111 (2007).
\bibitem{Fe92} H. Feshbach, {\it Theoretical Nuclear Physics} 
 Vol. II (Wiley, New York, 1992).
\bibitem{Br97} M.E. Brandan and G.R. Satchler, Phys. Rep. {\bf 285}, 143 (1997).
\bibitem{Be77} G. Bertsch, J. Borysowicz, H. McManus, and W.G. Love,
 Nucl. Phys. A {\bf 284}, 399 (1977).
\bibitem{An83} N. Anantaraman, H. Toki, and G.F. Bertsch,
 Nucl. Phys. A {\bf 398}, 269 (1983). 
\bibitem{Ko82} A.M. Kobos, B.A. Brown, P.E. Hodgson, G.R. Satchler, and 
A. Budzanowski, Nucl. Phys. A {\bf 384}, 65 (1982).
\bibitem{Kho93} D.T. Khoa and W. von Oertzen, Phys. Lett. B {\bf 304}, 8 (1993).
\bibitem{Kho95s} D.T. Khoa, G.R. Satchler, and W. von Oertzen,
 Phys. Rev. C {\bf 51}, 2069 (1995).
\bibitem{Brown} G.E. Brown, Rev. Mod. Phys. {\bf 43}, 1 (1971).
\bibitem{Migdal} A.B. Migdal, {\it Theory of Finite Fermi Systems and Applications 
 to Atomic Nuclei} (Interscience, New York, 1967).
\bibitem{HvH} N.M. Hugenholtz and L. Van Hove, Physica {\bf 24}, 363 (1958).
\bibitem{Cze02} P. Czerski, A. De Pace, and A. Molinari, 
 Phys. Rev. C {\bf 65}, 044317 (2002). 
\bibitem{Loa15} D.T. Loan, B.M. Loc, and D.T. Khoa, 
 Phys. Rev. C {\bf 92}, 034304 (2015).
\bibitem{Bru59} K.A. Brueckner and D.T. Goldman, Phys. Rev. {\bf 116}, 424 (1959).
\bibitem{Vau72} D. Vautherin and D.M. Brink, Phys. Rev. C {\bf 5}, 626 (1972).
\bibitem{Hee12} P.H. Heenen and M.R. Godefroid,  {\it The Hartree-Fock method}, 
 Scholarpedia 7 (10) 10545 (2012); http://www.scholarpedia.org/.
\bibitem{Hs75} P.E. Hodgson, Rep. Prog. Phys. {\bf 38}, 847 (1975).
\bibitem{Kho94} D.T. Khoa, W. von Oertzen, and H.G. Bohlen, 
  Phys. Rev. C {\bf 49}, 1652 (1994).
\bibitem{Kho01} D.T. Khoa, Phys. Rev. C {\bf 63}, 034007 (2001).
\bibitem{Kho96} D.T. Khoa, W. von Oertzen, and A.A. Ogloblin, Nucl. Phys. A {\bf 602}, 98 (1996).
\bibitem{Mah85} C. Mahaux, P.F. Bortignon, R.A. Broglia, and C.H. Dasso,
 Phys. Rep. {\bf 120}, 1 (1985).
\bibitem{Zuo99} W. Zuo, I. Bombaci, and U. Lombardo, Phys. Rev. C {\bf 60}, 024605 (1999).
\bibitem{Ma91} C. Mahaux and R. Sartor, Adv. Nucl. Phys. {\bf 20}, 1 (1991). 
\bibitem{BM69} A. Bohr and B.R. Mottelson, {\it Nuclear Structure}
 vol.~I, p.~237 (W.A. Benjamin, Inc., New York, 1969).
\bibitem{Va91} R.L. Varner, W.J. Thompson, T.L. McAbee, E.J. Ludwig, 
and T.B. Clegg, Phys. Rep. {\bf 201}, 57 (1991).
\bibitem{Hama} S. Hama, B.C. Clark, E.D. Cooper, H.S. Sherif, and R.L. Mercer,
  Phys. Rev. C {\bf 41}, 2737 (1990).
\bibitem{Fri88} S.H. Fricke, M.E. Brandan, and K.W. McVoy,
  Phys. Rev. C {\bf 38}, 682 (1988).
\bibitem{Bra96} M.E. Brandan, M.S. Hussein, K.W. McVoy, and G.R. Satchler, 
 Comments Nucl. Part. Phys. {\bf 22}, 77 (1996). 
\bibitem{Bri77} D.M. Brink and N. Takigawa, Nucl. Phys. A {\bf 279}, 159 (1977).
\bibitem{Row77} N. Rowley, H. Doubre, and C. Marty, Phys. Lett. B {\bf 69}, 147 (1977).
\bibitem{Bra88} M.E. Brandan and G.R. Satchler, Nucl. Phys. A {\bf 487}, 477 (1988).
\bibitem{Kho00O} D.T. Khoa, W. von Oertzen, H.G. Bohlen, and F. Nuoffer, 
 Nucl. Phys. A {\bf 672}, 387 (2000).
\bibitem{Oglob00} A.A. Ogloblin, Yu. A. Glukhov, W.H. Trzaska, A.S. Dem'yanova, 
 S.A. Goncharov, R. Julin, S.V. Klebnikov, M. Mutterer, M.V. Rozhkov, V.P. Rudakov,
 G.P. Tiorin, D.T. Khoa, and G.R. Satchler, Phys. Rev. C {\bf 62}, 044601 (2000).
\bibitem{Wie76} R.M. Wieland, R.G. Stokstad, G.R. Satchler, 
 and L.D. Rickertsen, Phys. Rev. Lett. {\bf 37}, 1458 (1976).
\bibitem{Sto79} R.G. Stokstad, R.M. Wieland, G.R. Satchler, C.B. Fulmer, D.C. Hensley, 
 S. Raman, L.D. Rickertsen, A.H. Snell, and P.H. Stelson, Phys. Rev. C {\bf 20}, 655 (1979).
\bibitem{McVoy92} K.W. McVoy and M.E. Brandan,  Nucl. Phys. A {\bf 542}, 295 (1992). 
\bibitem{Pol76} J.E. Poling, E. Norbeck, and R.R. Carlson, Phys. Rev. C {\bf 13}, 648 (1976).
\bibitem{Raynal} J. Raynal, in {\it Computing as a Language of Physics}
 (IAEA, Vienna, 1972) p.75;  J. Raynal, coupled-channel code ECIS97 (unpublished).
\bibitem{Hos88} J.Y. Hostachy, M. Buenerd, J. Chauvin, D. Lebrun, Ph. Martin, 
 J.C. Lugol, L. Papineau, P. Roussel, N. Alamanos, J. Arvieux, and C. Cerruti, 
 Nucl. Phys. A {\bf 490}, 441 (1988).
\bibitem{Kubono83} S. Kubono, K. Morita, M.H. Tanaka, M. Sugitani, H. Utsunomiya, 
 H. Yonehara, M.K. Tanaka, S. Shimoura, E. Takada, M. Fukuda, and K. Takimoto, 
 Phys. Lett. B {\bf 127}, 19 (1983).
\bibitem{Bohlen85} H.G. Bohlen, X.S. Chen, J.G. Cramer, P. Fr\"obrich,
 B. Gebauer, H. Lettau, A. Miczaika, W. von Oertzen, R. Ulrich, and Th. Wilpert, 
 Z. Phys. A {\bf 322}, 241 (1985). 
\bibitem{Alla10} A.S. Demyanova, H.G. Bohlen, A.N. Danilov, S.A. Goncharov, 
 S.V. Khlebnikov, V.A. Maslov, Yu.E. Penionzkevich, Yu.G. Sobolev, W. Trzaska, 
 G.P. Tyurin, and A.A. Ogloblin, Nucl. Phys. A {\bf 834}, 473c (2010).
\bibitem{Cole81} A.J. Cole, W.D.M. Rae, M.E. Brandan, A. Dacal, B.G. Harvey, 
 R. Legrain, M.J. Murphy, and R. G. Stokstad, Phys. Rev. Lett. {\bf 47}, 1705 (1981).
\bibitem{Buenerd84} M. Buenerd, A. Lounis, J. Chauvin, D. Lebrun, P. Martin, 
 G. Duhamel, J.C. Gondrand, and P. De Saintignon, Nucl. Phys. A {\bf 424}, 313 (1984).
\bibitem{Oglob03} A.A. Ogloblin, S.A. Goncharov, Yu. A. Glukhov, A.S. Demyanova, 
 M.V. Rozhkov, V.P. Rudakov, and W.H. Trzaska, Phys. Atomic Nuclei {\bf 66}, 1478 (2003).
\bibitem{Ful75} R.C. Fuller, Phys. Rev. C {\bf 12}, 1561 (1975).
\bibitem{Kondo98} Y. Kond\=o, M.E. Brandan, and G.R. Satchler, 
 Nucl. Phys. A {\bf 637}, 175 (1998). 
\bibitem{Bra91} M.E. Brandan and G.R. Satchler, Phys. Lett. B {\bf 256}, 311 (1991).
\bibitem{Oglob98} A.A. Ogloblin, D.T. Khoa, Y. Kond\=o, Yu.A. Glukhov, 
 A.S. Dem’yanova, M.V. Rozhkov, G.R. Satchler, and S.A. Goncharov, 
 Phys. Rev. C {\bf 57}, 1797 (1998). 
\bibitem{Demya} A.S. Dem’yanova {\it et al.}, IAEA Database EXFOR;\\  
 http://www-nds.iaea.org/exfor/.
\bibitem{Bra01} M.E. Brandan, A. Menchaca-Rocha, L. Trache, H.L. Clark,
 A. Azhari, C.A. Gagliardi, Y.-W. Lui, R.E. Tribble, R.L. Varner,
 J.R. Beene, and G.R. Satchler, Nucl. Phys. A {\bf 688}, 659 (2001). 
\bibitem{Bra86} M.E. Brandan, A. Menchaca-Rocha, M. Buenerd, J. Chauvin, 
 P. De Saintignon, G. Duhamel, D. Lebrum, P. Martin, G. Perrin, and J.Y. Hostachy, 
 Phys. Rev. C {\bf 34}, 1484 (1986). 
\bibitem{Rou88} P. Roussel-Chomaz, N. Alamanos, F. Auger, J. Barrette, 
 B. Berthier, B. Fernandez, and L. Papineau, Nucl. Phys. A {\bf 477}, 345 (1988).
\bibitem{vOe02} S. Szilner, W. von Oertzen, Z. Basrak, F. Haas, and M. Milin,
 Eur. Phys. J. A {\bf 13}, 273 (2002). 
\bibitem{Mor11} M.C. Morais and R. Lichtenth\"aler, Nucl. Phys. A {\bf 857}, 1 (2011). 
\bibitem{Rud10} A.T. Rudchik {\it et al.}, Eur. Phys. J. A {\bf 44}, 221 (2010).
\bibitem{Ohkubo14} S. Ohkubo and Y. Hirabayashi, 
 Phys. Rev. C {\bf 89}, 051601(R) (2014).
\bibitem{Ohkubo14i} S. Ohkubo, Y. Hirabayashi, A.A. Ogloblin, Yu.A. Gloukhov,
 A. S. Dem’yanova, and W.H. Trzaska, Phys. Rev. C {\bf 90}, 064617 (2014).
\bibitem{Minomo14} K. Minomo, M. Toyokawa, M. Kohno, and M. Yahiro, 
 Phys. Rev. C {\bf 90}, 051601(R) (2014).
\end{thebibliography}
\end{document}